\documentclass[reprint,superscriptaddress, amsmath,amssymb, aps, pra, ]{revtex4-2}

\usepackage[usenames, dvipsnames]{color}
\usepackage{comment}
\usepackage{hyperref}
\hypersetup{
    colorlinks=true,
    linkcolor=Blue,
    citecolor=Blue,
    filecolor=Blue,      
    urlcolor=Blue,
    pdftitle={Spinor droplets},
    }
\usepackage{graphicx}
\usepackage{physics}

\newcommand{\PsiCreate}[1]{\hat{\Psi}_{#1}^{\dagger}(\mathbf{x})}
\newcommand{\PsiAnn}[1]{\hat{\Psi}_{#1}(\mathbf{x})}
\newcommand{\akm}[2]{\hat{a}_{\mathbf{#1},#2}}
\newcommand{\akmDagger}[2]{\hat{a}^{\dagger}_{\mathbf{#1},#2}}
\newcommand{\FExpected}{\langle \mathbf{F}\rangle}
\newcommand{\Fmm}[1]{\mathbf{F}_{#1}}
\newcommand{\DkDagger}[1]{\hat{D}_{\mathbf{#1}}^\dagger}
\newcommand{\Dk}[1]{\hat{D}_{\mathbf{#1}}}
\newcommand{\FkDagger}[1]{\mathbf{\hat{F}}_{\mathbf{#1}}^\dagger}
\newcommand{\Fk}[1]{\mathbf{\hat{F}}_{\mathbf{#1}}}
\newcommand{\ek}{\epsilon_{\mathbf{k}}}
\newcommand{\cbar}[1]{c_{#1}}
\newcommand\scalemath[2]{\scalebox{#1}{\mbox{\ensuremath{\displaystyle #2}}}}

\def\xBold{\mathbf{x}}
\def\U0{\tilde{U}(0)}

\def\a-kd{a^{\dagger}_{-k}}

\def\b-kd{b^{\dagger}_{-k}}
\def\epsk{\epsilon_{\mathbf{k}}}

\def\sumk0{\sum_{\vec{k} \neq 0}}
\def\sqrtbeta{\sqrt{\beta_2^2 + 2\beta_0-z_4}}
\def\omegaTilde{\tilde{\omega}}

\begin{document}

\title{Polarized Rabi-Coupled and Spinor Boson Droplets }

\author{T.A. Yo\u{g}urt}
 \email{ayogurt@metu.edu.tr}
 \affiliation{%
Department of Physics, Middle East Technical University, Ankara, 06800, Turkey\\
}%
\author{A. Kele\c{s}}
 \affiliation{%
Department of Physics, Middle East Technical University, Ankara, 06800, Turkey\\
}%

\author{M.\"O. Oktel}
\affiliation{Department of Physics, Bilkent University, Ankara, 06800, Turkey}

\date{\today}

\begin{abstract}
Self-bound quantum droplets form when the mean-field tendency of the gas to collapse is stabilized by the effectively repulsive beyond mean-field fluctuations.  The beyond mean-field effects depend on Rabi-frequency $\omega_R$ and quadratic Zeeman effect $q$ for the Rabi-coupled Bose mixtures and the spinor gases, respectively. The effects of varying $\omega_R$ and $q$ on the quantum droplet have recently been examined for unpolarized Rabi-coupled Bose mixture with zero detuning $\delta = 0$ and unpolarized spinor gas with $\langle F_z \rangle=0$.  In this paper, we theoretically explore the stability of the droplet phase for polarized $\delta \neq 0$ Rabi-coupled Bose mixture and $\langle F_z \rangle \neq 0$ spinor gas. We calculate the Lee-Huang-Yang corrections for both gases with polarized order parameters and obtain the phase diagram of the droplets on the parameter space of $\omega_R$-$\delta$ and $q$-$p$ for Rabi-coupled mixture and spinor gas, respectively. Finally, we highlight the similarities and differences between the two systems and discuss their experimental feasibility.    
\end{abstract}

\maketitle

\section{\label{sec:level1} INTRODUCTION}

Theoretical prediction and experimental realization of the bosonic droplets strikingly highlight the significance of the beyond mean-field (MF) effects which generally give minor corrections.  Self-trapping of a Bose-Einstein condensate (BEC) that is otherwise collapsing is only possible if the beyond MF fluctuations are taken into account \cite{2016_PRA_Santos_Dipolar_Droplet_Theory,2015_Petrov_PRL}. In a trapped single component BEC with attractive interactions, the MF interaction energy scales with $-N^2|a|/R^3$, where $N$ is the number of particles, $R$ is the radius of the condensate, and $a$ is the $s$-wave scattering length ($a<0$). The trapping potential $\propto NR^2$ and kinetic energy $\propto N/R^2$ of the condensate may balance this attractive MF interaction and yield a metastable BEC only if the particle number is below some maximum value \cite{pethick2008bose}. The situation is drastically different in the case of self-bound droplets. There, the MF collapse is stabilized by the beyond mean-field (BMF) quantum fluctuations even without a confining potential 
\cite{2015_Petrov_PRL,
2016_PRA_Santos_Dipolar_Droplet_Theory,
2016_PRL_Pfau_Dipolar_Droplet,
2016_Nature_Pfau,
2016_PRL_Mazzanti,
2016_PRA_Blakie_Dipolar_Droplet,
2017_Nature_Salasnisch_RabiCoupled,
2018_Science_Tarruel_Mixture_Droplet,
2018_RPL_Modugno_Mixture_Droplet,
2019_PRA_Oktel,
2020_PRA_Mazzanti_SpinOrbitDroplet,
2021_PRA_Cornish_Mixture_Droplet,
2021_Arxiv_Cui_Borromean_Droplet,
2021_PRL_Santos_Dipolar_Mixture,
2022_PRA_Yogurt_Spinor}, and 
droplets exhibit a minimum particle number below which the gas is no longer stable \cite{2015_Petrov_PRL}. As particle number $N$ decreases,the kinetic energy eventually dominates and causes the gas to expand.

In addition to these constraints in the particle number, the gas is also required to be in 
the dilute regime in which the Bogoliubov theory is still valid to achieve droplet formation. Therefore, stability of droplets depends on the interaction parameters. Consider the three different classes of Bose droplets: dipolar \cite{2016_PRA_Santos_Dipolar_Droplet_Theory}, binary mixture \cite{2015_Petrov_PRL}, and spinor \cite{2022_PRA_Yogurt_Spinor}. For the dipolar droplets, the dipole $\epsilon_{dd}$ and contact interactions $a_s$; for binary mixtures, intraspecies $a_{11}$ and $a_{22}$ and interspecies $a_{12}$ contact interactions; for spin-1 gas, the total spin-$0$ channel $a_0$ and total spin-$2$ channel $a_2$ interactions are the fundamental interaction parameters to be considered. These parameters should be fine-tuned to drive the MF energy of the system towards collapse and balance the collapse with the BMF energy. Requirement of such stringent fine-tuning motivates a search for additional probes to adjust droplet formation in cold atom experiments. 

Recently, the Bose mixtures with the Rabi-coupling between the hyperfine states of the particles attracted attention due to interesting many-body effects, such as effective tunable three-body interactions \cite{2017_Nature_Salasnisch_RabiCoupled,2019_PRA_Chiquillo_Rabicoupled_lowdimensional,2021_PRL_Bourdel_Rabicoupled_BMF,2022_PRL_Bourdel_Rabicoupled_Threebody}. The coupling between the two levels of the system makes the BMF energy depend on the Rabi frequency $\omega_R$ \cite{2021_PRL_Bourdel_Rabicoupled_BMF} and provides an additional mechanism to tune the droplet density. Similar to a critical particle number, there is a critical Rabi frequency $\omega_c$ above which the droplet is no longer self-trapped \cite{2017_Nature_Salasnisch_RabiCoupled}. For the spinor droplets, the quadratic Zeeman energy-dependent LHY correction of the spinor gas plays an analogous role with a maximum quadratic Zeeman energy $q_c$ above which the droplet expands \cite{2022_PRA_Yogurt_Spinor}. Additionally, when non-zero quadratic Zeeman energy or Rabi-frequency is introduced, one of the gapless Bogoliubov modes becomes gapped for both spinor and Rabi-coupled mixtures. The MF energies can be controlled by the detuning $\delta$ for Rabi-coupled gases and linear Zeeman energy $p$ for spinor gases. These similarities prompt us to investigate their droplet states comparatively.  

Previously, both the Rabi-coupled binary mixture droplets \cite{2017_Nature_Salasnisch_RabiCoupled} and  the spinor gas droplets \cite{2022_PRA_Yogurt_Spinor} are studied for zero net polarization. For the Rabi-coupled droplets, the detuning $\delta$ is assumed to be zero, which yields an unpolarized ground state order parameter within the MF picture. Similarly, for the spinor droplet \cite{2022_PRA_Yogurt_Spinor}, the MF ground states is studied with zero magnetization $\langle F_z \rangle$. 
In this paper, we theoretically explore the droplet formation for the polarized Rabi-coupled binary mixture and the spin-1 gas. We examine how the non-zero polarization affects the MF and BMF energies and discuss the feasibility of the droplet phases under finite polarization. For the Rabi-coupled binary mixtures, the non-zero detuning $\delta$ leads to an asymmetry in the particle number of the two levels within the MF ground state. Finite polarization alters both the MF and BMF interaction energies. For a given  Rabi-frequency $\omega_R$, there is a critical value of the detuning $\delta_c$ above which the droplet is not self-trapping. Similarly, for the spin-1 droplet, finite magnetization alters the MF and BMF interactions and a critical magnetization exists $\tilde{p}_c$ for given quadratic Zeeman energy $q$. 
Using these critical values, we obtain the droplet phase boundary of the Rabi-coupled mixture and spinor gas in the $\omega_R$-$\delta$ and $q$-$p$ planes, respectively. 

This paper is organized as follows. In Section \ref{sec:Bogoliubov Theory of Rabi-Coupled Bose Mixture}, we summarize the Bogoliubov Theory of Rabi-coupled binary mixtures and discuss the possible MF ground states and mechanical stability of the mixture. In Section \ref{sec:Rabi-Coupled Bose Mixture Droplet}, we develop the formulation of the polarized Rabi-coupled droplet, present our numerical results on the droplet phase boundary in the $\omega_R$-$\delta$ plane. In Section \ref{sec:Bogoliubov Theory of Spin-1 Gas}, we summarize the Bogoliubov Theory of polarized spin-1 gas and discuss the mean-field order parameters for anti-ferromagnetic interactions $c_1>0$. In Section \ref{sec:Spin-1 Droplet}, we develop the formulation of the polarized spinor droplet and present  the droplet phase boundary in the $p-q$ plane. In Section \ref{sec:Experimental Discussion and Conclusion}, we discuss the experimental feasibility of the proposed phenomena and highlight the similarities and differences between the polarized Rabi-coupled and spinor droplets. 

\section{\label{sec:Bogoliubov Theory of Rabi-Coupled Bose Mixture} Rabi-Coupled Bose Mixtures: Bogoliubov Theory}
We consider a BEC consisting of $N$ atoms in  two internal states, $m = 1, 2$ with the corresponding $s$-wave scattering lengths
$a_{11}$, $a_{22}$ and $a_{12}$. The internal states are coupled through a Rabi frequency $\omega_R$ and detuning $\delta$. Applying the rotating wave approximation to eliminate the explicit time dependence, the Hamiltonian of this Rabi-coupled binary mixture is given by \cite{2021_PRL_Bourdel_Rabicoupled_BMF}:
\begin{eqnarray}
\hat{H} &&= \int d\xBold{} \ \biggl\{ \sum_{m=1,2} \PsiCreate{m} \left( -\frac{\hbar^2 \nabla^2}{2M} \right) \PsiAnn{m} \nonumber \\ 
&&+ \sum_{m,m'} \left( \frac{g_{mm'}}{2} \ \PsiCreate{m} \PsiCreate{m'} \PsiAnn{m'} \PsiAnn{m} \right) \nonumber \\ 
&& -\hbar \omega_R\left(\PsiCreate{1}\PsiAnn{2} + \PsiCreate{2}\PsiAnn{1}\right) \nonumber \\ 
&&-\hbar \frac{\delta}{2}\left( \PsiCreate{2}\PsiAnn{2}-\PsiCreate{1}\PsiAnn{1}\right) \biggl\}  
\label{RabiCoupledHamiltonian}
\end{eqnarray}
where $g_{mm'} =4\pi a_{mm'} \hbar^2/M$ are the coupling constant of the $s$-wave interaction among the atoms of mass $M$ within the internal states $m$ and $m'$. $\PsiCreate{m}$ and $\PsiAnn{m}$ are the field operators that create and annihilate the particle with internal state $m$ at position $\xBold{}$, respectively. 

We obtain the MF energy and BMF fluctuations using the Bogoliubov theory in the Hamiltonian \eqref{RabiCoupledHamiltonian}. Assuming a homogenous gas, we express the field operators in terms of Fourier modes $\PsiAnn{m} = V^{-1/2}\sum_{\mathbf{k}} \akm{k}{m} e^{i\mathbf{k}\mathbf{x}}$, and write the operators $\akm{k}{m} = \akm{0}{m} + \sum_{\mathbf{k}\neq 0} \akm{k}{m}$ keeping only the terms up to the quadratic order in  $\akm{\mathbf{k}\neq 0}{m}$. We replace the operators with the classical number $\akm{0}{m} \approx \sqrt{N_{0,m}}$, where $N_{0,m}$ is the number of particles with internal state $m$ in the $\mathbf{k} = 0$ state. The Hamiltonian \eqref{RabiCoupledHamiltonian} becomes: 
\begin{eqnarray}
\hat{H} &&= - 2\hbar \omega_R\sqrt{N_1N_2}-\frac{\hbar \delta}{2} (N_2-N_1) + \sum_{m,m'} \frac{g_{mm'}N_m N_{m'}}{2V} \nonumber\\
&&+  \sum_{\mathbf{k} \neq 0} \Biggl\{ \left(\epsilon_{\mathbf{k}}  + g_{11}n_1 +  \hbar \omega_R \sqrt{\frac{N_2}{N_1}} \right)\akmDagger{k}{1}\akm{k}{1} \nonumber \\
&&+ \left(\epsilon_{\mathbf{k}}  + g_{22}n_2 +  \hbar \omega_R \sqrt{\frac{N_1}{N_2}} \right)\akmDagger{k}{2}\akm{k}{2} \nonumber \\
&&+ \frac{g_{11}n_1}{2}\left( \akmDagger{k}{1}\akmDagger{-k}{1} + \akm{k}{1}\akm{-k}{1}\right)\nonumber \\ 
&&+ \frac{g_{22}n_2}{2}\left( \akmDagger{k}{2}\akmDagger{-k}{2} + \akm{k}{2}\akm{-k}{2}\right)\nonumber \\ 
&&+ g_{12}\sqrt{n_1n_2}\left( \akmDagger{k}{1}\akmDagger{-k}{2} + \akm{k}{1}\akm{-k}{2}\right)\nonumber \\
&&+ \left(g_{12}\sqrt{n_1n_2}-\hbar \omega_R \right)\left( \akmDagger{k}{1}\akm{k}{2} + \akmDagger{k}{2}\akm{k}{1}\right) \Biggl\}
\label{RabiCoupledHamiltonianinmomentum}
\end{eqnarray}
where $\epsilon_{\mathbf{k}} = \frac{\hbar^2 k^2}{2M}$ is the free particle dispersion. The first line of \eqref{RabiCoupledHamiltonianinmomentum} is the MF energy of the Rabi-coupled binary mixture gas, while the rest of the terms account for the quantum fluctuations that constitute the BMF energy. 
\begin{figure}
    \centering
    \includegraphics{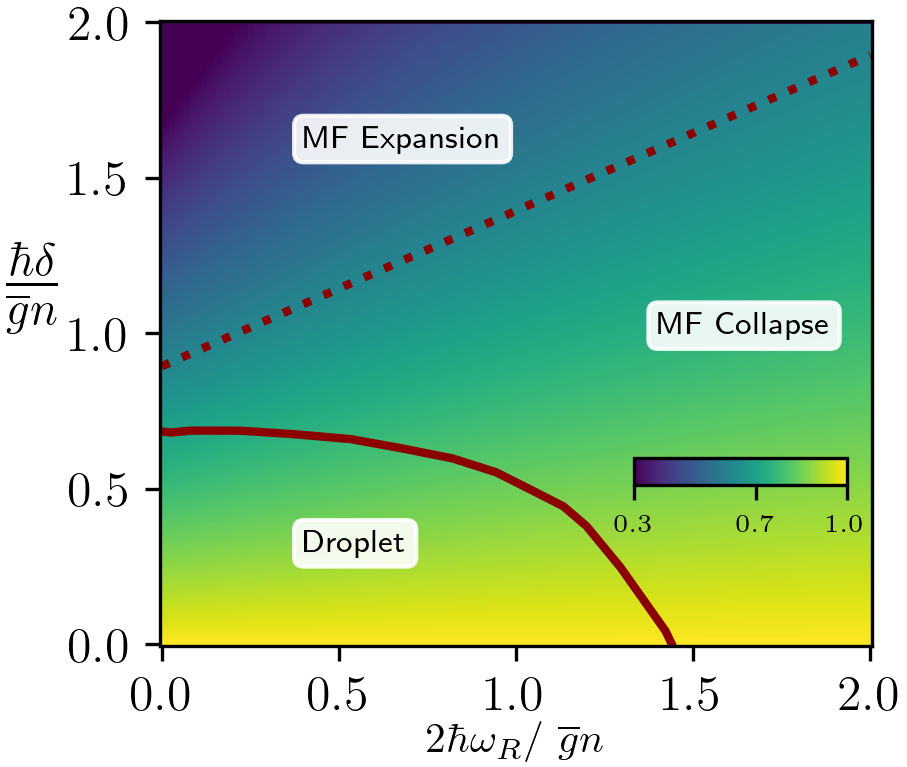}
    \caption{ \label{fig:Rabicoupled_MF_Phase_Diagram} Mean-field phase diagram of Rabi-coupled Bose mixture as a function of Rabi-frequency $\frac{\hbar \omega_R}{\bar{g}n}$ and detuning $\frac{\hbar \delta}{\bar{g}n}$ for $\frac{g_{12}}{g}<-1$, and $g>0$. The shade (color bar) indicates the ratio $r = \sqrt{N_1/N_2}$ in the MF ground state \eqref{RabicoupledMFEnergyFunctional}. The dashed (red) line indicates the boundary above or below which the total MF two-particle interaction 
    forces the gas to expand or collapse, respectively, which is drawn for the scattering length ratio of $a_{12}/a = -1.5$   The solid (red) line is the boundary of the self-trapped droplet phase.}
\end{figure}

In order to understand the ground state order parameter within the MF picture and how it differs from the binary mixture without Rabi-coupling, we focus on the MF energy and assume that intra-species scattering lengths are equal for both internal states, i.e. $g_{11} = g_{22} = g$ for simplicity. The MF energy from \eqref{RabiCoupledHamiltonianinmomentum} is given by \cite{2022_PRL_Bourdel_Rabicoupled_Threebody}: 
\begin{eqnarray}
\frac{E_{MF}}{N} = -\hbar \omega_R \sin{\theta} - \frac{\hbar \delta}{2} \cos{\theta} + \frac{gN}{2V} - \frac{\bar{g} N}{2V} \sin^2{\theta}
\label{RabicoupledMFEnergyFunctional}
\end{eqnarray}
where $\bar{g} = \frac{g-g_{12}}{2}$ and the wavefunctions of the condensate components $(\psi_1 \ \psi_2) = \sqrt{n} (\sin{\theta/2}\ \cos{\theta/2})$ with $\theta \in [0,\pi]$.  Assuming the total number of particles $N$ fixed, the problem of determining the MF ground state becomes finding $\theta$ that minimizes the energy \eqref{RabicoupledMFEnergyFunctional}. The ground state MF order parameter $r \equiv \sqrt{N_1/N_2} = \tan (\theta/2)$ for any $g_{12}/g$ with $\delta = 0$ can be found in \cite{2017_Nature_Salasnisch_RabiCoupled}. Here, we are interested in the parameter space for $\delta \neq 0$. Since our purpose is to examine the order parameters in which the system can collapse within the MF picture, we assume $g_{12}/g<-1$ and $g>0$, which gives collapse without phase separation. 
The ratio between the particle numbers $r$ within the MF ground state for various $\omega_R$ and $\delta$ values are shown in Fig.~\ref{fig:Rabicoupled_MF_Phase_Diagram}.
For any $\omega_R$ with $\delta = 0$, the MF energy is minimized by $\theta = \pi/2$ or $r=1$.
This is exactly the order parameter to which Salasnich \textit{et al.} \cite{2017_Nature_Salasnisch_RabiCoupled} restrict  their droplet analysis. As detuning $\delta$ becomes non-zero, the MF ground state becomes polarized $r \neq 1$. The polarization becomes sharper, i.e. $r \rightarrow 0$ or $\theta \rightarrow 0$, as either $\omega_R \rightarrow 0$ or $\delta \rightarrow \infty$.  The order parameter $r = \tan(\theta/2)$, and $\theta \in [0,\pi/2]$ changes smoothly over the parameter space (See Fig.\ref{fig:Rabicoupled_MF_Phase_Diagram}).

Now let us discuss how the two-body interaction part ($\frac{gN}{2V} - \frac{\bar{g} N}{2V} \sin^2{\theta}$) of the MF energy (\ref{RabicoupledMFEnergyFunctional}) changes with the detuning $\delta$. On the $\omega_R$-axis, the MF ground state yields $r=1$ or $\theta = \pi/2$ which gives the two-body interaction $\propto (g-\bar{g})n^2 = (g+g_{12})n^2/2$. Since $g/g_{12}<-1$, the density collapse is expected within the MF picture. However, on the $\delta$-axis, the MF ground state yields $r=0$ or $\theta = 0$ for $\hbar \delta/\bar{g}n >2$. Hence, the MF two-body interaction is $\propto gn^2$. Since $g>0$, the gas expansion is expected within the MF picture. Note that as detuning $\delta$ is increased from zero to infinity, the MF order parameter $\theta$ changes from $\pi/2$ and approaches to zero. As a consequence, we expect a value of $\theta = \sin^{-1}{\sqrt{g/\bar{g}}}$, below or above which the MF picture predicts a collapse or expansion, respectively. For $g_{12}/g = -1.5$, the line that separates the density collapse and expansion within the MF picture is shown with the dashed red line in Fig.\ref{fig:Rabicoupled_MF_Phase_Diagram}. 

Below, the free parameters of the Rabi-coupled mixture will be taken as $\omega_R$, $r$, $g_{12}$, $g$, and $N$. Furthermore, the dimensionless parameters $\tilde{\omega} = \frac{\hbar \omega_R}{gn}$ and $\gamma = g_{12}/g$ will be used when appropriate. The results will be presented as functions of parameter set ($\tilde{\omega}$,$r$) which then can be mapped to the parameter plane ($\omega_R$,$\delta$) when necessary.

We calculate the BMF energy to analyse the possibility of droplet phase for various values of $\omega_R$ and $r$. By applying a Bogoliubov transformation on the quadratic Hamiltonian \eqref{RabiCoupledHamiltonianinmomentum}, one can obtain the Bogoluibov modes of the Rabi-coupled binary mixture \cite{2021_PRL_Bourdel_Rabicoupled_BMF}:
\begin{widetext}
\begin{eqnarray}
E_{\pm,\mathbf{k}}&=& \sqrt{D_{\mathbf{k}} \pm \sqrt{D_{\mathbf{k}}^2-\epsilon_{\mathbf{k}}\left( \epsilon_{\mathbf{k}}+\hbar \omega_R \left(r + \frac{1}{r}\right)\right)\left[ \left(\epsk{} + 2gn_1 + \hbar \omega_R \frac{1}{r} \right)\left(\epsk{} + 2gn_2 + \hbar \omega_R r \right)- \left( 2g_{12}\sqrt{n_1n_2}-\hbar \omega_R\right)^2\right]}}  \nonumber\\
D_{\mathbf{k}} &=& \frac{1}{2} \sum_{m=1,2} \left[ \left( \epsk{} + \hbar \omega_R \sqrt{\frac{n_{\bar{m}}}{n_m}} \right)\left(\epsk{} + 2gn_m + \hbar \omega_R \sqrt{\frac{n_{\bar{m}}}{n_m}} \right)- \hbar \omega_R (2g_{12}\sqrt{n_1n_2}-\hbar \omega_R)\right]
\label{Rabicoupled_Bog_Modes}
\end{eqnarray}
\end{widetext}
where 
$\bar{m}=3-m$, 
$n_1 = \frac{r^2}{r^2 + 1}n$ and $n_2 = \frac{1}{r^2+1}n$. These Bogoliubov modes reduce to the results of Salasnisch \textit{et al.} \cite{2017_Nature_Salasnisch_RabiCoupled} for $r=1$ and non-zero $\omega_R$ and they recover the usual Bose mixture results without Rabi coupling for $\omega_R = 0$ and $r=1$.  
 
We calculate the BMF energy of each corresponding Bogoluibov mode separately using $E^{\pm}_{BMF} = \frac{1}{2} \sum_{\mathbf{k}} \left( E^{\pm}_{\mathbf{k}}- \lim_{\mathbf{k} \rightarrow \infty} E^{\pm}_{\mathbf{k}}\right)$, which gives:
\begin{equation}
    \frac{E_{BMF}^{\pm}}{V} = \alpha (gn)^{5/2} I_{\pm}(\tilde{\omega}, \gamma, r)
    \label{Rabicoupled_LHY_Energy}
\end{equation}
where $\alpha = \frac{m^{3/2}}{\sqrt{2}\pi^2\hbar^3}$,
and we neglect one of the modes containing imaginary part. See Appendix (\ref{sec:I_pm_Expression}) for $I_{\pm}(\tilde{\omega}, \gamma, r)$.

In the limit $\tilde{\omega}\rightarrow0$ and $r=1$, this result recovers the BMF energy of the Bosonic mixture without Rabi-coupling. The limit $r\rightarrow1$ for any $\tilde{\omega}$, the $E_{BMF}^{\pm}$ expressions of Salasnich \textit{et.al.} \cite{2017_Nature_Salasnisch_RabiCoupled} are recovered for both modes.
For more general cases ($r \neq 1$), we calculate the $I_+$ numerically for various $r$ values.
For any $r\in [0.6,1]$, $I_+$ is a monotonically increasing function of both $\tilde{\omega}$ and $r$ so that BMF energy of the Rabi-coupled mixture increases with these parameters.

\section{\label{sec:Rabi-Coupled Bose Mixture Droplet} Rabi-Coupled Bose Mixture Droplet}

\begin{figure*}[t]
    \includegraphics[width=0.4\textwidth]{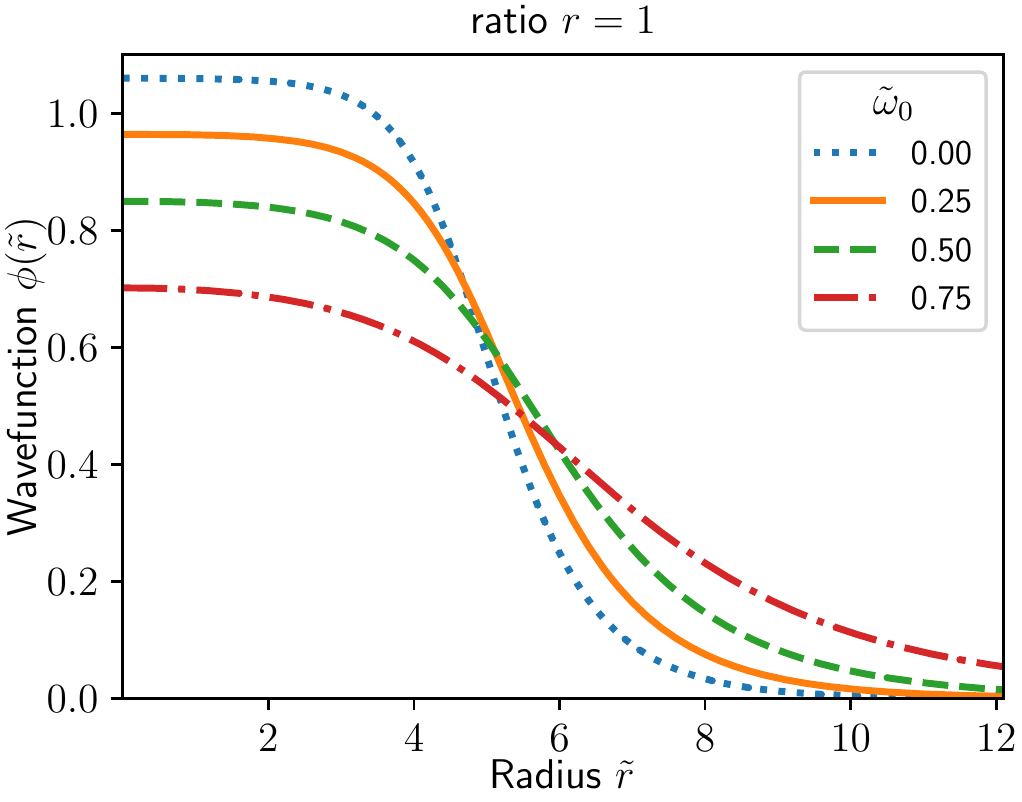}
    \includegraphics[width=0.4\textwidth]{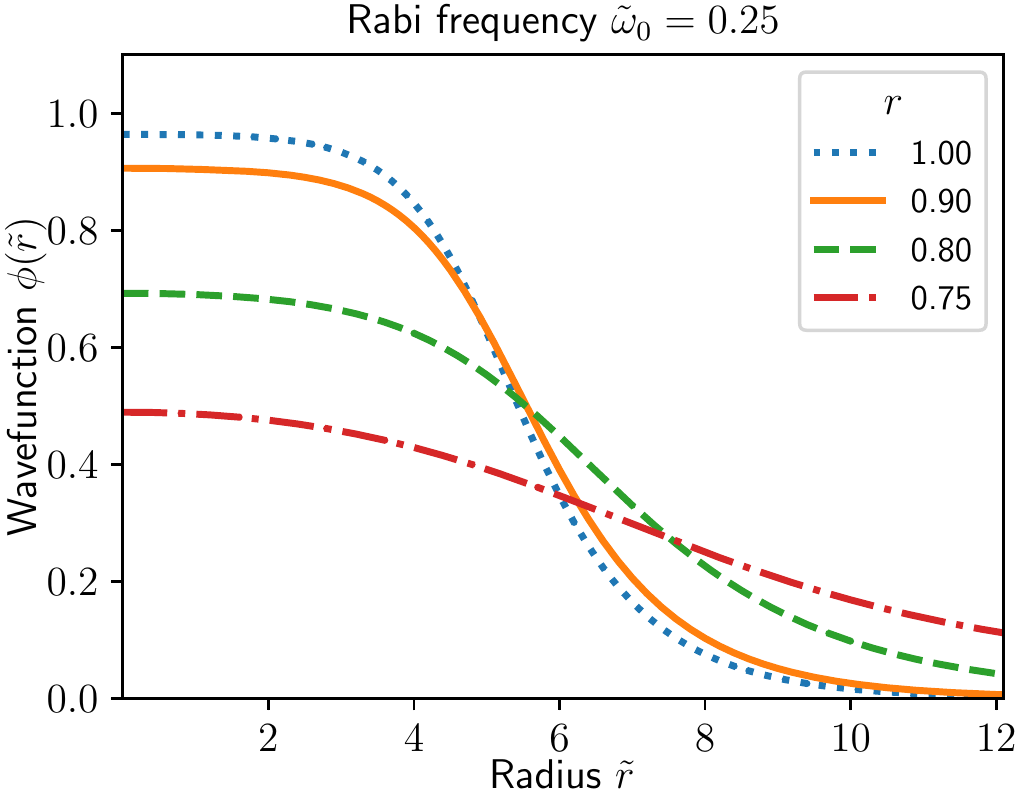}
    \caption{ \label{RC_Droplet_Wavefunctions} The ground state wavefunctions of the Rabi-coupled Bose mixture in polar phase for various values of $\tilde{\omega}$ and ratio $r$. (Left) The wavefunctions for $\delta = 0$, i.e. $r=1$ and different values of  $\tilde{\omega}$. Above $\tilde{\omega}_c$ droplet is no longer self trapped. (Right) The wavefunctions for fixed $\tilde{\omega} = 0.25$ and varying ratio $r$, which shows self-bound droplet until a critical value of $r_c \approx 0.7$. The total particle number $\tilde{N}=500$ for both plots. } 
\end{figure*}
We now discuss the possibility of self-trapping and neglect the `soft' Bogoliubov mode contribution $E_{\mathbf{k}}^{-}$, as in Refs. \cite{2015_Petrov_PRL,2019_PRA_Oktel,2022_PRA_Yogurt_Spinor}.
We first consider the infinite, homogeneous Rabi coupled Bose mixture. The pressure of the gas is calculated from $P = -\partial \left( E_{MF} + E_{BMF}^{+} \right)/\partial V$ as follows:
\begin{eqnarray}
P = \frac{g(1+r^4) + 2g_{12}r^2}{2(r^2+1)^2}n^2 
+ \alpha  \left(gn\right)^{5/2} f(\tilde{\omega})
\end{eqnarray}
where $f(\tilde{\omega}) = \frac{3}{2}I_+(\tilde{\omega})-\tilde{\omega}I_+^{'} (\tilde{\omega})$. For any $g_{12}/g$, there is a value of $r\in [0,1]$ above which the pressure due to the MF energy is negative. Furthermore, this negative pressure can be stabilized by a positive contribution from BMF energy, since $g>0$ and $f(\tilde{\omega})$ is positive for any value of $\tilde{\omega}$. Under these circumstances, the vanishing pressure $P=0$ condition can be reached. We obtain an implicit equation for the equilibrium density: 
\begin{eqnarray}
n_0 = \frac{\left[ g(1+r^4) + 2g_{12}r^2\right]^2}{4(r^2+1)^4  \alpha^2 g^5 f^2(\tilde{\omega}_0)}
\end{eqnarray}
where $\tilde{\omega}_0 = \frac{\hbar \omega_R}{gn_0}$. If $\omega_R = 0$ and $r=1$, this equilibrium density becomes $n_0^{(1)} = \frac{25 |\delta g|^2}{16  \alpha^2 g^5 (1+|\gamma|)^5}$, where $\delta g \equiv g_{12} + g$. Here, $n_0^{(1)}$ also approximates the density of the finite droplets in which the kinetic energy is negligible. As the Rabi-frequency $\tilde{\omega}$ is increased for a fixed ratio $r$, the function $f(\tilde{\omega})$ and BMF energy becomes greater, which in turn decreases the equilibrium density of the droplet.

We study the feasibility of the finite droplet more quantitatively by obtaining the governing Gross-Pitaevskii Equation (GPE). We use locked-in approximation for the different components of the mixture and assume a droplet wavefunction $\Psi(\mathbf{r}) = \psi(\mathbf{r}) (\tau_1 \ \tau_2)^T$, where $\tau_1/\tau_2=r$ and $|\tau_1|^2 + |\tau_2|^2 = 1$. We express the energy functional of the droplet using $n(\mathbf{r}) = |\Psi(\mathbf{r})|^2$ as
\begin{eqnarray}
\mathcal{E}[\psi^*,\psi] &=& \frac{\hbar^2}{2M}|\nabla\psi|^2 
+ \left( -\frac{2\hbar \omega_R r }{r^2 + 1 }-\frac{\hbar \delta (1-r^2)}{2(1+r^2)}\right) |\psi|^2 
\nonumber\\
&+& \left( \frac{g}{2}-\frac{2\bar{g}r^2}{(1+r^2)^2}\right) |\psi|^4
\nonumber\\
&+& \alpha  g^{5/2} I_{+}\left(\frac{\hbar \omega_R}{g|\psi|^2}, \gamma, r\right) |\psi|^5
\end{eqnarray}
and write the wavefunction in dimensionless form $\psi(\mathbf{r}) = \sqrt{n_0^{(1)}} \phi(\mathbf{r}) $. We minimize the total energy in the grand canonical ensemble $E=\int d^3\mathbf{r}\mathcal{E}[\psi^*,\psi]-\mu N$ where the chemical potential is fixed by the total number of particles $N=\int d^3\mathbf{r}|\psi|^2$. The resulting modified GPE is given by:
\begin{eqnarray}
\tilde{\mu} \phi &=& -\frac{1}{8} \left( r + \frac{1}{r} \right)^2 \tilde{\nabla}^2 \phi \nonumber \\
&+& \Biggl[ 2\alpha_4 |\phi|^2 + \frac{5\alpha_5}{2} I_+\left(\frac{\tilde{\omega}_0^{(1)}}{|\phi|^2},r,\gamma \right) |\phi|^3 
\nonumber \\
&-& \alpha_5 \tilde{\omega}_0^{(1)} I^{'}_{+} \left( \frac{\tilde{\omega}_0^{(1)}}{|\phi|^2},r,\gamma \right) | \phi | \Biggl] \phi 
\label{GP_Equation_RC_Mixture}
\end{eqnarray}
where $\tilde{\omega}_0^{(1)} = \frac{\hbar \omega_R}{gn_0^{(1)}}$, $\alpha_4 = \frac{3}{2|\delta g|} \left( \frac{(r^2+1)^2g}{2r^2 }-2\bar{g}\right)$, and $\alpha_5 = \frac{15}{8(1 + |\gamma|)^{5/2}}\left( r + \frac{1}{r}\right)^2$. The equation \eqref{GP_Equation_RC_Mixture} is written in the dimensionless form $\tilde{\mathbf{r}}=\mathbf{r}/\xi$, with $\xi = \sqrt{\frac{6\hbar^2}{M |\delta g|n_0^{(1)}}}$ and the total particle number is scaled by $\tilde{N} = N/n_0^{(1)}\xi^3$. This modified GPE reduces to the form obtained by Petrov \cite{2015_Petrov_PRL} when $\omega_R = 0$ and $r=1$, as $\alpha_4 = -3/2$ and $\alpha_5 I_+(0,1,\gamma) = 1$ in this limit. Below, we fix the scattering length ratio $\gamma = -1.5$,  as in Salasnich \textit{et al.} \cite{2017_Nature_Salasnisch_RabiCoupled}.

We numerically solve the modified GPE \eqref{GP_Equation_RC_Mixture} by imaginary time evolution and obtain the ground state wavefunction. For a fixed total particle number $\tilde{N} = 500$, we find the critical value of $\tilde{\omega}$ above which the droplet expands to infinity. This expansion occurs with a mechanism different from the expansion due to decreasing total particle number $\tilde{N}_c$. In the latter case, the decreasing number of particles causes the droplet radius to shrink so that the increasing kinetic energy causes expansion to infinity. In the case of increasing $\tilde{\omega}$ at fixed ratio $r$, the droplet radius increases due to a stronger BMF energy which makes kinetic energy comparable to both MF and BMF energies, and causes the droplet to expand. As an example, for $\tilde{N}=500$ and $\delta = 0$, or $r=1$, 
the critical frequency is $\tilde{\omega}_c = 0.9$. Fig.~(\ref{RC_Droplet_Wavefunctions}) shows the increase in the droplet radius with increasing $\tilde{\omega}$ up to the critical $\tilde{\omega}_c$ in the left panel. As $r$ decreases, the MF interaction energy ($\alpha_4$ term in Eq.~\ref{RabicoupledMFEnergyFunctional}) first decreases then acts repulsive for $r<0.62$. BMF energy \eqref{Rabicoupled_LHY_Energy} also decreases with $r$ but since MF energy shrinks with a higher rate, we expect $\tilde{\omega}_c$ to become smaller with decreasing $r$.

As $r$ changes from $1$ to $0.7$, $\tilde{\omega}_c$ changes from $0.9$ to $0$, and we do not observe a droplet phase below $r=0.7$. We show the droplet wavefunction for various ratio $r$ for fixed $\tilde{N}=500$ and $\omegaTilde{} = 0.5$ on the right panel of Fig.(\ref{RC_Droplet_Wavefunctions}).

We numerically obtain the critical $\tilde{\omega}_c$ values for different $r$ and fixed $\tilde{N}= 500$ to obtain the boundary of the droplet phase as shown in Fig.(\ref{fig:Rabicoupled_MF_Phase_Diagram}).

\section{\label{sec:Bogoliubov Theory of Spin-1 Gas} Spin-1 Gases: Bogoliubov Theory}
In previous work \cite{2022_PRA_Yogurt_Spinor}, we studied the spin-1 gas with vanishing magnetization $\langle F_z \rangle = 0$ and found that the spinor droplet is possible in the polar and antiferromagnetic phases if density interaction is negative $c_0<0$ and spin interaction is positive $c_1>0$. The quadratic Zeeman energy $q$ in spinor gas is analogous to the Rabi-frequency $\omega_R$ in Bose mixtures and both can tune the density of the droplet. As $q$ increases, the BMF energy causes the droplet to expand and beyond a critical level of $q$, the gas cannot self-bind. Similarly, the detuning $\delta$ in the Rabi-coupled mixture is analogous to the linear Zeeman shift  $p$ in the spinor gas. 

\begin{figure}[t]
    \includegraphics[scale=1.2]{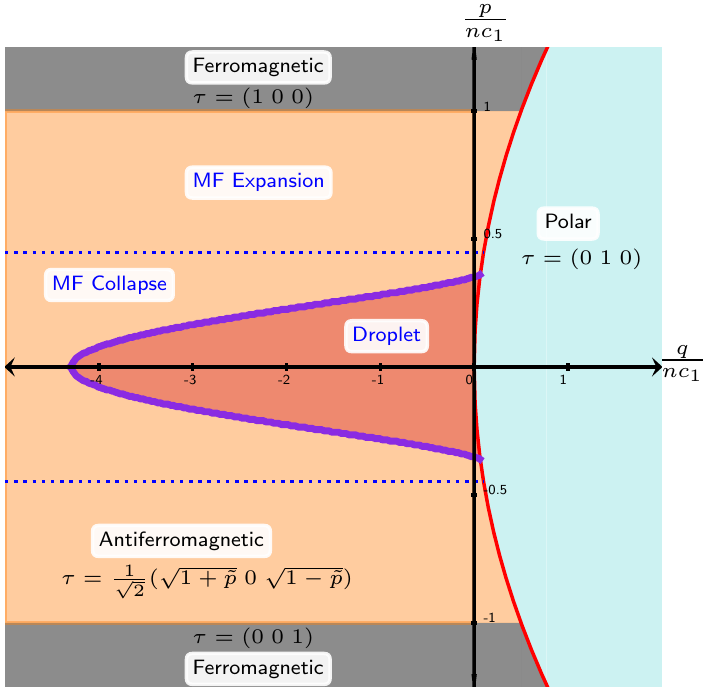}
    \caption{ \label{fig:spinor_MF_phase_diagram} Mean-field phase diagram of spin-1 gas as a function of quadratic  $q/nc_1$ and linear $p/nc_1$ Zeeman energies. The orange region corresponds to the antiferromagnetic order $\tau^{AF} = 1/\sqrt{2} (\sqrt{1+\tilde{p}}\ 0 \ \sqrt{1-\tilde{p}})$. MF theory predicts an expansion of the gas outside the dashed (blue) lines $|\tilde{p}|>0.44$, and density collapse inside $|\tilde{p}|<0.44$. The droplet phase boundary is shown with solid (purple) line where the gas can be stabilized by BMF fluctuations. The total particle number $\tilde{N}=500$ and $c_1/c_0 = -5$ with $c_0<0$.}  
\end{figure}

Here, we extend our spinor droplet discussion to include the effects of non-zero magnetization $p \neq 0$. The ground state order parameter changes only for the anti-ferromagnetic phase 
(See Fig.(\ref{fig:spinor_MF_phase_diagram})) and it gives a constant shift in the MF energy for polar phase \cite{2012_PR_Ueda_Spinor_BEC,2013_RMP_Spinor_Kurn_Ueda,2011_PRA_Kurn_Spinor}.  

Spin-1 BEC with s-wave interactions and a uniform magnetic field along the $z$-axis is described by the following Hamiltonian: 
\begin{eqnarray}
\hat{H} &&= \int d\xBold{} \ \Biggl\{ \PsiCreate{m} \left( -\frac{\hbar^2 \nabla^2}{2M} + qm^2 -pm \right)\PsiAnn{m}\nonumber \\
&& + \frac{\cbar{0}}{2} \ \PsiCreate{m} \PsiCreate{m'} \PsiAnn{m'} \PsiAnn{m} \\&&+ \frac{\cbar{1}}{2} \  \PsiCreate{m} \PsiCreate{m'}\ \Fmm{m n} \cdot \ \Fmm{m' n '} \PsiAnn{n '} \PsiAnn{n}  \Biggl\} \nonumber 
\label{spin1Hamiltonian}
\end{eqnarray}
where  $\PsiCreate{m}$ and $\PsiAnn{m}$ create and annihilate spin-1 atom in the magnetic quantum state $m=-1,0,1$,
$\Fmm{mm'} = (F_{mm'}^x,F_{mm'}^y,F_{mm'}^z)$ are the spin-1 matrices in $z$-axis basis, and summation convention is used for $m$ indices, $p = -g_L\mu_B B_{eff}$ is the product of Land\'e $g_L$-factor, the Bohr magneton $\mu_B$ and the effective magnetic field $B_{eff}$.
The quadratic Zeeman energy $q = q_B + q_{MW}$ can be tuned using both an external static field $q_B = \frac{(g\mu_B B)^2}{\Delta E_{hf}}$ and microwave field $q_{MW}$. Interactions in the density and spin channels are parametrized by coupling constants $c_0$ and $c_1$. With the Bogoliubov approximation, this Hamiltonian reduces to quadratic form 
\cite{2010_PRA_Ueda_Spinor_LHY}: 
\begin{eqnarray} 
\hat{H}_{eff} = &&\frac{Vn^2}{2}(c_0 + c_1 \FExpected^2) + qN\langle F_z^2 \rangle -pN\langle F_z \rangle \nonumber \\
&& + \sum_{\mathbf{k \neq 0}}  \Big \{ \left[ \ek - nc_1 \FExpected^2 + qm^2 - q\langle F_z^2 \rangle \right] \akmDagger{k}{m} \akm{k}{m} \nonumber\\
&&+ nc_1\FExpected \cdot \Fmm{mm'}  \akmDagger{k}{m} \akm{k}{m'} \nonumber\\ &&+ \frac{nc_0}{2}(2\DkDagger{k} \Dk{k} + \Dk{k}\Dk{-k} + \DkDagger{k} \DkDagger{-k}) \nonumber \\
&&+ \frac{nc_1}{2}(2\FkDagger{k} \Fk{k} + \Fk{k}\Fk{-k} + \FkDagger{k} \FkDagger{-k}) \Big \}  
\label{spin1QuadraticHamiltonian}
\end{eqnarray}
where $\ek = \hbar^2 \mathbf{k}^2/2M$ is the free particle dispersion, $\FExpected \equiv \sum_{m,m'} \Fmm{mm'}\tau_m^*\tau_{m'}$ is the expectation value of the spin-1 order parameter, $\Dk{k} \equiv \sum_{m} \tau_m^* \akm{k}{m}$ and $\Fk{k} \equiv \sum_{m,m'} \Fmm{mm'} \tau_m^* \akm{k}{m'}$ are the density and spin fluctuation operators, $N_0$ is the number of particles in the $\mathbf{k} = 0$ state, $\tau$ is the ground state order parameter in the spin-1 manifold. 

The MF energy of spin-1 BEC obtained from  \eqref{spin1QuadraticHamiltonian} is
\begin{eqnarray}
\frac{E_\mathrm{MF}}{V} = &&\frac{n^2}{2}(c_0 + c_1 \FExpected^2) + qn\langle F_z^2 \rangle -pn\langle F_z \rangle 
\label{MF_energy_Spinor}
\end{eqnarray}
whereas all the other terms within the summation constitute the quantum fluctuations. 
The MF ground state order parameter $\tau$ is determined by minimizing the energy \eqref{MF_energy_Spinor}. We consider the magnetic orders when the spin coupling constant is positive $c_1>0$ as shown in Fig.(\ref{fig:spinor_MF_phase_diagram}). For $p = 0$, the order parameter is $\tau_P = (0 \ 1 \ 0)$ if $q>0$, or $\tau_{AF} = 1/\sqrt{2}(1 \ 0 \ 1)$ if $q<0$. When $q>0$, introducing non-zero $p$ does not make any difference in $\tau_P$ around the $q$-axis. However, if $q<0$, the MF energy is minimized by a $p$ dependent order parameter $\tau_{AF} = 1/\sqrt{2}(\sqrt{1 + \tilde{p}}\ 0 \ \sqrt{1 - \tilde{p}})$, where $\tilde{p} \equiv p/nc_1$. Note that $\langle F_z\rangle = \tilde{p}$, hence, $\tau_{AF}$ is defined for $-1\le \tilde{p} \le 1$. Outside this interval, the order parameter becomes ferromagnetic. In this paper, we focus on this $\tilde{p}$ dependent $\tau_{AF}$ to analyze how the spinor droplets in the antiferromagnetic phase are affected when non-zero magnetization $\tilde{p}$ is introduced.

The order parameter $\tau_{AF}$ gives $\langle F_z^2 \rangle = 1$, $\langle F_z \rangle = \tilde{p}$, and $\langle \mathbf{F} \rangle = \tilde{p}\hat{e}_z$ and the resulting quadratic Hamiltonian is:
 
\begin{eqnarray}
&\hat{H}& = E_{0}^{AF}   + \sum_{\mathbf{k} \neq 0} \Biggl\{ \left(\epsilon_{\mathbf{k}}  -q + c_1n \right)\akmDagger{k}{0}\akm{k}{0}\nonumber \\
&+& \frac{c_1n\beta}{2} \left(\akmDagger{k}{0} \akmDagger{-k}{0} + \akm{k}{0}\akm{-k}{0}\right) \nonumber \\
&+& \sum_{m=\pm 1} \left[ \epsk{} + \frac{(c_0 + c_1)n(1+m\tilde{p})}{2}\right] \akmDagger{k}{m}\akm{k}{m} \nonumber \\
&+& \sum_{m=\pm 1} \left[ \frac{(c_0 + c_1)n(1+m\tilde{p})}{4}\right] \left( \akmDagger{k}{m}\akmDagger{-k}{m}+\akm{k}{m}\akm{-k}{m}\right) \nonumber \\
&+& \frac{(c_0-c_1)n\beta}{4} \biggl( 2\akmDagger{k}{-1}\akm{k}{1} + \akmDagger{k}{-1}\akmDagger{-k}{1} + \akmDagger{k}{1}\akmDagger{-k}{-1}  \nonumber \\
&+& 2\akmDagger{k}{1}\akm{k}{-1} + \akm{k}{-1} \akm{-k}{1} + \akm{k}{1} \akm{-k}{-1} \biggl)  \Biggl\} 
\label{Spin1HamiltonianQuadratic}
\end{eqnarray}
where $\beta = \sqrt{1-\tilde{p}^2}$, and $E_0^{AF}$ is the MF energy. Dispersion of three distict Bogoliubov modes are found as 
\begin{eqnarray}
&E&_{k,\pm1} = \sqrt{\epsk{\left[ \epsk{ + (c_0+c_1)n(1\pm \kappa) }\right]}} \\
&E&_{k,0} = \sqrt{\left( \epsk{}-q + (1-\beta)c_1n \right) \left(\epsk{}-q + (1+\beta)c_1n\right)}\nonumber
\label{Spinor_Bog_Modes}
\end{eqnarray}
where $\kappa \equiv \sqrt{1-\frac{4\beta^2c_0c_1}{(c_0+c_1)^2}}$. We add the renormalization terms for each mode by using the T-matrix approach up to the second order \cite{2010_PRA_Ueda_Spinor_LHY} and obtain the following energy density including both MF and BMF energy: 
\begin{eqnarray}
\frac{E_{0}}{V} &=& (q-\tilde{p}p)n + \frac{(c_0 + c_1 \tilde{p}^2)n^2}{2} \nonumber \\
&+& \frac{8\sqrt{2}}{15} \alpha (c_1n)^{5/2} I_0(\tilde{q},\beta) \nonumber \\
&+& \frac{8\sqrt{2}}{15} \alpha \left( (c_0+c_1)n\right)^{5/2} [I_{+}(\kappa) + I_{-}(\kappa)] 
\label{Spinor_BMF_Energies}
\end{eqnarray}
where $\tilde{q} = \frac{q}{nc_1}$, $I_{\pm} = \frac{(1\pm \kappa)^{5/2}}{4\sqrt{2}}$ and $I_0(\tilde{q},\beta)$ can be approximated as (See Appendix \ref{sec:I0_approximation}): 
\begin{eqnarray}
I_0(\tilde{q},\beta) \approx \frac{15 \pi \beta^2}{32 \sqrt{2}} \left[\sqrt{-\tilde{q} + 1} -\frac{\beta^2}{32}\frac{1}{(-\tilde{q}+1)^{3/2}} \right]
\label{I0_integral_approximation}
\end{eqnarray}

The first line in \eqref{Spinor_BMF_Energies} is the MF energy density for the order parameter $\tau^{AF}$ with non-zero $\tilde{p}$, the second and third lines are the BMF energy contributions from the three different Bogoliubov modes. This expression reproduces the results given in 
Ref.~\onlinecite{2022_PRA_Yogurt_Spinor} for $\tilde{p}=0$ and $\tau_{AF} = 1/\sqrt{2}\ (1 \ 0 \ 1)$.
 
Notice that when $\tilde{p}=0$, the MF density favors collapse if $c_0<0$. Interestingly, MF energy decreases in magnitude as $|\tilde{p}|$ increase, and if $\tilde{p}>\sqrt{|c_0|/c_1}$, it becomes repulsive, which leads an expansion of the gas above a critical level shown with dotted lines in Fig.\ref{fig:spinor_MF_phase_diagram}. 

\begin{figure*}[t]
\label{fig:spinor_droplet_wavefunctions}
    \includegraphics[width=0.4\textwidth]{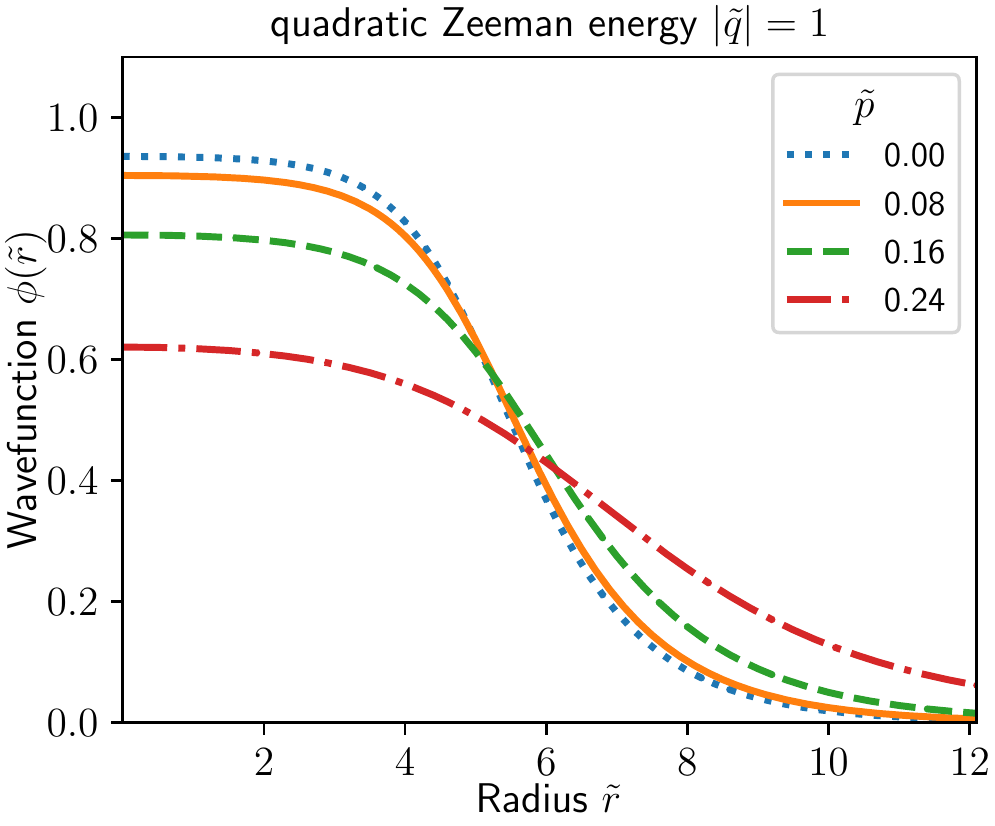}
    \includegraphics[width=0.4\textwidth]{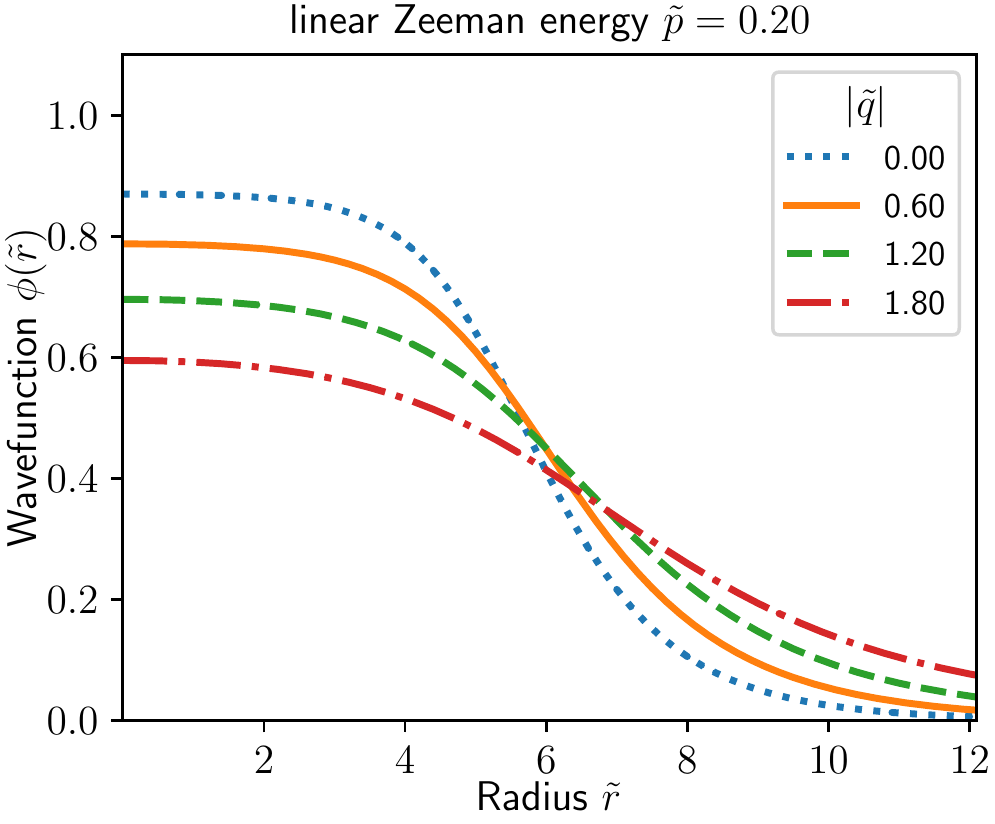}
    \caption{ \label{fig:Spinor_Droplet_Wavefunctions} The ground state wavefunctions of the spinor gas in AF phase for various values of quadratic Zeeman $\tilde{q}$ and linear Zeeman $\tilde{p}$ energy. The total particle number $\tilde{N}=500$ and $c_1/c_0=-5$ with $c_0<0$ for both plots. (Left) The wavefunctions for $|\tilde{q}| = 1$ and different values of  $\tilde{p}$. Above $\tilde{p}_c$ droplet is no longer self trapped. (Right) The wavefunctions for fixed $\tilde{p} = 0.2$ and varying $\tilde{q}$. Similarly, above $\tilde{q}_c$, self-bound droplet cannot be formed.} 
\end{figure*}

In the collapse regime $\tilde{p}<\sqrt{|c_0|/c_1}$, the contribution of the BMF energy is repulsive since $c_1>0$, and it can stabilize the gas. The hard modes given above by $E_+$ and $E_0$ dispersion provide such stabilization whereas the soft mode $E_-$ containing imaginary part can be neglected, similar to previous droplet studies \cite{2015_Petrov_PRL,2022_PRA_Yogurt_Spinor}.

\section{\label{sec:Spin-1 Droplet} Polarized AF Spin-1 Droplet}

In the parameter regime $c_0<0$ and $c_1>0$, the pressure of the gas is calculated using the thermodynamic identity $P=-\partial E/\partial V$ with the total energy given by $E=E_{MF} + E_{BMF}^{+} + E_{BMF}^{0}$ which gives:

\begin{eqnarray}
P = \left(\frac{c_0 + \tilde{p}^2c_1}{2}\right)n^2 + \frac{4\sqrt{2}}{15}\alpha (c_1n)^{5/2} \ h(\tilde{q},\beta)  
\label{Polarized_spinor_droplet_pressure}
\end{eqnarray}
where $h(\tilde{q},\beta) = 3I_0(\tilde{q},\beta) + 3\left(c_0/c_1+1\right)^{5/2}I_+(\kappa) - 2\tilde{q}I_0^{'}(\tilde{q},\beta)$. Here, prime on $I_0$ denotes the partial derivative with respect to $\tilde{q}$. The equilibrium density for the infinite homogeneous droplet can be found from the vanishing pressure %
\begin{eqnarray}
n_0 = \frac{225}{128}\frac{(c_0 + \tilde{p}^2c_1)^2}{c_1^5\  h^2(\tilde{q},\beta)}
\label{Equilibrium_density_polarized_spinor_droplet}
\end{eqnarray}
which is equivalent to the equilibrium density result of Ref.~\onlinecite{2022_PRA_Yogurt_Spinor} for zero magnetization $\tilde{p}=0$. We take the limit $q\rightarrow 0$ to obtain a density scale $n_0^{(1)} = \frac{25|c_0|^2}{512\alpha^2|c_1|^5}$ and use it to express the dimensionless modified GPE. Since $h(\tilde{q},\beta)$ is a monotonically increasing function of $\tilde{q}$, the equilibrium density decrease with increase of the quadratic Zeeman energy $\tilde{q}$. 
Larger $\tilde{q}$ provides stronger BMF fluctuations and the system can stabilize at lower densities.

We   use the locked-in approximation $\Psi(\mathbf{r}) = \psi(\mathbf{r}) \tau^{AF} $ with $\tau^{AF} = 1/\sqrt{2}\left( \sqrt{1 + \tilde{p}}\ 0 \ \sqrt{1 - \tilde{p}}\right)$ and write the energy functional
\begin{eqnarray}
&\mathcal{E}&[\psi^*,\psi] = \frac{\hbar^2}{2M}|\nabla\psi|^2 \nonumber \\&+& \left( q-\tilde{p}p\right) |\psi|^2+ \left( \frac{c_0 + c_1\tilde{p}^2}{2}\right) |\psi|^4
\\
&+&  \frac{8\sqrt{2}}{15}\alpha \left[ (c_1n)^{5/2} I_0(\tilde{q},\beta) + \left((c_0 + c_1)n\right)^{5/2}I_+(\kappa)\right]|\psi|^5. \nonumber
\end{eqnarray}
Using $\psi(\mathbf{r}) = \sqrt{n_0^{(1)}} \phi(\mathbf{r})$, we minimize the total energy $E=\int d^3\mathbf{r}\mathcal{E}[\psi^*,\psi]-\mu N$ with the total number of particles $N=\int d^3\mathbf{r}|\psi|^2$ which yields the modified GPE
\begin{eqnarray}
\label{GP_Equation_Spinor}
\tilde{\mu} \phi &=& -\frac{1}{2} \tilde{\nabla}^2 \phi \\
&+& \Biggl\{ -3\left(1 + \frac{c_1}{c_0}\tilde{p}^2 \right) |\phi|^2 + \frac{5}{4} I_0\left(\frac{\tilde{q}_0}{|\phi|^2},\beta\right)|\phi|^3 \nonumber \\&+& \frac{5}{4}\left( \frac{c_0}{c_1}+1\right)^{5/2}I_+(\kappa)|\phi|^3 \nonumber - \frac{\tilde{q}_0}{2} I_0^{'}\left(\frac{\tilde{q}_0}{|\phi|^2},\beta\right) |\phi|\Biggl\} \phi  \nonumber
\end{eqnarray}
where $\tilde{q}_0 = \frac{q}{n_0^{(1)}c_1}$,
$\mathbf{\tilde{r}} = \mathbf{r}/\xi$, $\xi = \sqrt{6\hbar^2 /M|c_0|n_0^{(1)}}$. 
In the limit $\tilde{p}\rightarrow 0$, we recover the GPE of the unpolarized AF gas $\tau^{AF} = 1/\sqrt{2}(1 \ 0 \ 1)$ with $I_+(\kappa) \rightarrow \left(\frac{c_1}{c_0 + c_1}\right)^{5/2}$, which  is expected to give a droplet phase up to a critical $|\tilde{q}|\approx 4.4$ for $\tilde{N} \approx 500$ and $c_1/c_0 = -5$ \cite{2022_PRA_Yogurt_Spinor}. 

When $|\tilde{p}| \ge \sqrt{c_1/|c_0|} \approx 0.45$, blue dashed line in Fig.(\ref{fig:spinor_MF_phase_diagram}), MF interaction becomes repulsive, the gas goes through expansion and BMF fluctuations provide corrections for further repulsion. 
When $|\tilde{p}| < \sqrt{c_1/|c_0|}$ the MF drives a density collapse while BMF interactions are still effectively repulsive. Typically, effect of $\tilde p$ is much more pronounced in the MF terms than the BMF corrections whereas effect of $|\tilde{q}|$ is small in MF interactions, but it strengthens the BMF fluctuations for given $\tilde{p}$. 

We display the droplet wavefunctions obtained from the numerical solution of modified GPE for $\tilde{q}=1$ with varying $\tilde{p}$ on the left, and for $\tilde{p}= 0.2$ with varying $\tilde{q}$ on the right panel of Fig.(\ref{fig:Spinor_Droplet_Wavefunctions}). 
For a fixed $\tilde{p}$, larger $\tilde{q}$ gives stronger BMF repulsion, which widens the droplet radius. For $\tilde{p}=0.2$, after $\tilde{q}_c \approx -2.2$, the gas cannot form a droplet. For a fixed $\tilde{q}$, greater magnetization means both lower MF attraction and lower BMF repulsion. But even a small BMF repulsion is sufficient to expand the gas, since the MF attraction becomes much weaker. After a critical level of $\tilde{p}_c \approx 0.3$ for  $\tilde{q}=1$, the gas cannot bind into a droplet. We obtain the critical levels for each $\tilde{q}$ and $\tilde{p}$ within the parameter region of interest and show the droplet phase boundary in Fig.(\ref{fig:spinor_MF_phase_diagram}) with the solid red curve.

\section{\label{sec:Experimental Discussion and Conclusion} Discussion of experimental realization and Conclusion}

The parameters of the phase diagrams discussed above are within current experimental capabilities for the Rabi coupled gas. Consider a mixture of $^{39}K$ atoms in the hyperfine states $|F=1,m_F = 0\rangle$ and $|F=1,m_F = -1\rangle$. The Feshbach resonance around $B \approx 54.5 \ G$ can be used to tune the intracomponent scattering lengths as $a_{11} = a_{22} = 40a_B$ and the intercomponent scattering length $a_{12} = -60a_B$ \cite{2017_Nature_Salasnisch_RabiCoupled}, where $a_B$ is the Bohr radius. The ratio of interactions  give  $\gamma = \frac{g}{g_{12}} = -1.5$. In the absence of detuning and Rabi coupling, $N \approx 23,000$ particles gives a droplet of radius $0.4 \ \mu m$ with a peak density $n_0 = 4.12 \times 10^{16} cm^{-3} $. For zero detuning $\delta = 0$, one can introduce a Rabi-coupling $\omega_R = 2\pi f_R$ up to the level $f_R \approx 51 \ kHz$. As the Rabi coupling frequency increases, the droplet expands to a radius $r \approx 0.65 \ \mu m$ and the density at the center of the droplet decreases to $n_0 \approx 0.8 \times 10^{16} cm^{-3}$. Above $51 \ kHz$, the droplet will not be self-bound. The role of non-zero detuning can be tested by setting the Rabi-frequency to $f_R = 10.2 \ kHz$ for the same number of particles. The critical level of detuning for these parameters is $\delta_c = 41 \ kHz$, which give density $n_0 = 3.75 \times 10^{16} cm^{-3} $ and radius $7.4 \ \mu m$ beyond which it is no longer self bound. 

Experimentally realized spinor BECs so far are not favorable for obtaining a spinor droplet since they are all mechanically stable $c_0>0$ \cite{1998_Nature_Ketterle_Spinor,2004_PRL_Chapman_spinor,2013_RMP_Spinor_Kurn_Ueda,2007_PRL_Let_RF_Quadratic,2020_PRR_Choi}.  While the use of Feshbach resonance is not possible,  the spinor BEC scattering lengths may be tuned using theoretically proposed optical Feshbach resonances in future cold atom settings \cite{2015_PRA_Optical_Feschbach_Resonance,2018_Nature_Ott_Optical_Feshbach_Resonance}. The scattering lengths that favor droplet formation can be estimated
considering  an atom with scattering lengths $a_0 = -50a_B$ in spin-0 channel and $a_2 = 20a_B$ in spin-1 channel with Land\'e factor $g_L = 1/2$ ($s = 1/2, l = 0, I = 3/2$) which gives $c_1/c_0 = -5$ with $c_0<0$. For zero linear and quadratic Zeeman energies, the spinor droplet with density $8.3 \times 10^{16} \ cm^{-3}$ and radius $0.6 \ \mu m$ can be formed with total particle number $N \approx 130,000$. This droplet will be self-bound until a critical level of quadratic Zeeman energy $q \approx 680 \ kHz$. For an initial magnetization per particle $\tilde{p} = 0.2$, the gas will be stable until the quadratic Zeeman energy exceeds $320 \ kHz$ where the density of the droplet at its center will be around $2.1 \times 10^{16} \ cm^{-3}$ and the radius $0.95 \  \mu m$.

In conclusion,  Rabi-coupled Bose mixture and spinor gas are similar to each other in the following ways: (i) The BMF energies are Rabi-frequency or quadratic Zeeman energy dependent, (ii) one of the Bogoluibov modes become gapped when non-zero Rabi-frequency or quadratic Zeeman energy is introduced, and (iii) the polarization, hence the effective mean-field energy, can be significantly changed using the detuning or linear Zeeman energy. Therefore, droplet formation and its properties are highly affected by the linear and quadratic Zeeman energies in spinor gases, the Rabi-frequency, and the detuning in the Bosonic mixtures. 

\appendix
\section{The integral expression for $I_{\pm}(\tilde{\omega},r,\gamma)$}
\label{sec:I_pm_Expression}
\begin{figure}
     \centering
     \includegraphics[width=0.5\textwidth]{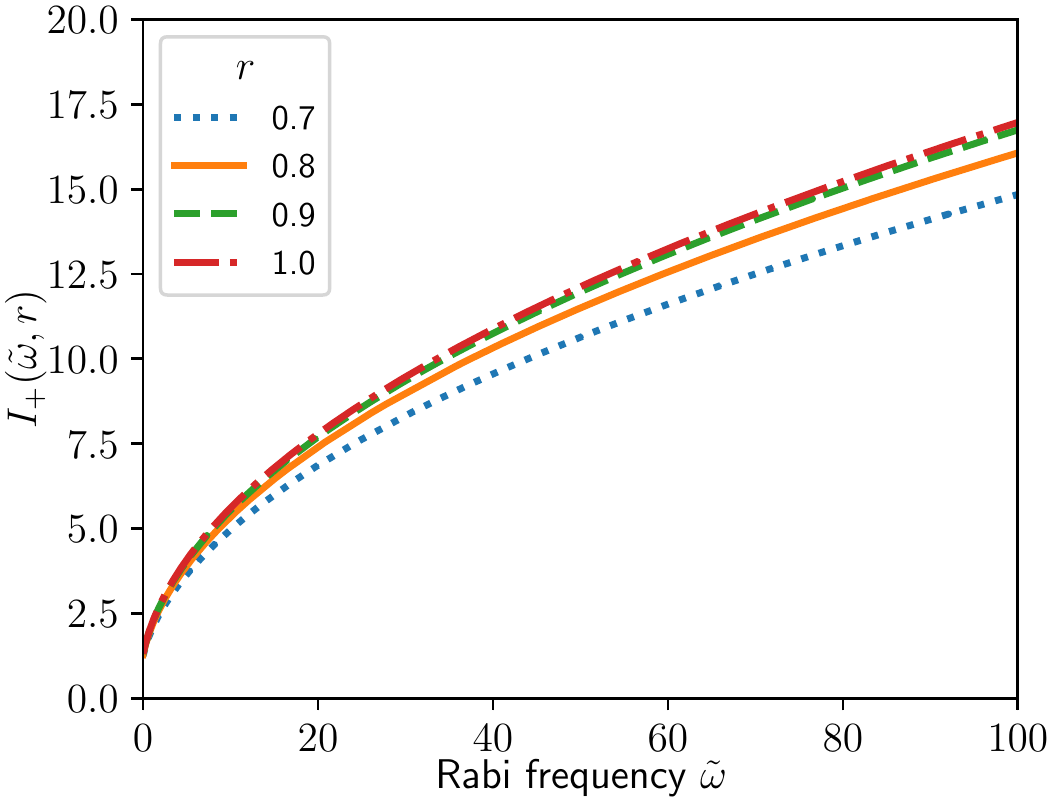}
     \caption{The integral $I_{+}$ in \eqref{Rabicoupled_I_pm_integrals} as a function $\tilde{\omega}$ for various $r$ and fixed $g_{12}/g = -1.5$. }
     \label{fig:I_plus_Integral_for_different_r}
\end{figure}

The integral expression for the function $I_{\pm}(\tilde{\omega},r,\gamma)$ within the BMF energy (\ref{Rabicoupled_LHY_Energy}) is given by:  
\begin{widetext}
\begin{eqnarray}
I_{\pm}(\tilde{\omega}, \gamma, r) \equiv \left(\frac{r^2}{r^2 + 1}\right)^{5/2} \int_0^{\infty} dy  y^2 & \left\{  \sqrt{(y^4 + \beta_2y^2 + \beta_0) \pm \sqrt{(\beta_2^2 + 2\beta_0 -z_4)y^4 + (2\beta_0\beta_2-z_6)y^2 + \beta_0^2}} \right. \nonumber\\
 &- y^2 \left. \left[ 1 + \frac{\beta_2 \pm \sqrtbeta{}}{2y^2} + \frac{\beta_0 \pm \frac{\beta_0\beta_2-z_6/2}{\sqrtbeta{}}-\frac{(\beta_2\pm \sqrtbeta{})^2}{4}}{2y^4}\right] \right\} 
 \label{Rabicoupled_I_pm_integrals}
 \end{eqnarray}
where
\begin{eqnarray}
\scalemath{1}{\beta_0 \equiv \frac{(r^2+1)^2}{r^4}\left[\frac{2r\omegaTilde{}}{r^2+1}(1-\gamma) + \omegaTilde{}^2 + \frac{\omegaTilde{}^2}{2r^2} + \frac{\omegaTilde{}^2r^2}{2}\right]}  \\
\scalemath{1}{\beta_2 \equiv \left(1 + \frac{1}{r^2}\right) \left( 1 + \frac{\tilde{\omega}}{r} + \tilde{\omega}r \right)} \\
\scalemath{1}{z_4 \equiv \left( 2 + \frac{\omegaTilde{}(r^2+1)}{r^3}\right) \left( \frac{2}{r^2} + \frac{\omegaTilde{}(r^2+1)}{r}\right) + \omegaTilde{} \frac{(r^2+1)^3}{r^5}\left(2 + \frac{\omegaTilde{}(r^2 +1)}{r} \right)-\left(\frac{2\gamma}{r}-\frac{\omegaTilde{}(r^2+1)}{r^2}\right)^2} \\
\scalemath{1}{z_6 \equiv \omegaTilde{}\frac{(r^2+1)^2}{r^3} \left[ \left(2 + \frac{\omegaTilde{}(r^2+1)}{r^3}\right) \left( \frac{2}{r^2} + \frac{\omegaTilde{}(r^2+1)}{r}\right)-\left(\frac{2\gamma}{r}-\frac{\omegaTilde{}(r^2+1)}{r^2}\right)^2\right]}
\end{eqnarray}
\end{widetext}
where  $gn_1y^2 \equiv \epsk{} $, $\omegaTilde{} = \frac{\hbar \omega_R}{gn} $, the particle number ratio $r = \sqrt{\frac{N_1}{N_2}}$, and scattering length ratio $\gamma = \frac{g_{12}}{g}$. 
Check Fig.\ref{fig:I_plus_Integral_for_different_r} to see how $I_+(\tilde{\omega}, \gamma, r)$ behaves for various $r$ values as $\tilde{\omega}$ changes.

\section{Analytical approximation for $I_0(\tilde{q},\beta)$}
\label{sec:I0_approximation}
The integral that determines the LHY energy for  $E_{k,0}$ mode reads:
\begin{eqnarray}
 &&I_0(\tilde{q},\beta) = \frac{15}{8\sqrt{2}} \int_0^{\infty}dx \ x^2 \nonumber \\
 &\times& \left[ \sqrt{(x^2-\tilde{q} + 1)^2 -\beta^2} -(x^2 - \tilde{q} + 1) + \frac{\beta^2}{2x^2}\right]
 \label{Spinor_I0_integral}
\end{eqnarray}
where $\epsk{} \equiv c_1n x^2$ substitution is done.

We use a change of variable $y \equiv x^2 -\tilde{q} + 1$ in
the integral \eqref{Spinor_I0_integral} and expand $\sqrt{1-\beta^2/y^2}$ in Taylor series up to the second order 
in the domain $x\ge 0$ and $\tilde{q} \le 0$ and obtain
\begin{eqnarray}
I(t)  &&=  -\frac{15\beta^2}{16\sqrt 2} \int_0^{\infty} dx  \left(\frac{(\tilde{q}-1)}{x^2 + t +1} + \frac{\beta^2 x^2}{4(x^2 + t +1)^3}\right) \nonumber \\
\label{I1_numerical_Appendix3}
\end{eqnarray}
Each term above can be calculated to give:
\begin{eqnarray}
I(t)  \approx \frac{15\pi \beta^2}{32\sqrt{2}} \left[ \sqrt{-\tilde{q}+1} - \frac{\beta^2}{32} \frac{1}{(-\tilde{q}+1)^{3/2}} \right].
\label{I1_analytical_Appendix} 
\end{eqnarray}
Higher order terms in the expansion of $\sqrt{1-\beta^2/y^2}$ improves the accuracy but we numerically checked that the second order expansion is sufficient to achieve less than one percent error for all $\tilde{q}$ values.

\acknowledgments
After the completion of this work, we became aware of a recent study \cite{2022_Arxiv_Cui_Polarized_RabiCoupled} related to polarized Rabi-coupled Bose mixture.

This work is supported by TUBITAK 2236 Co-funded
Brain Circulation Scheme 2 (CoCirculation2) Project No.
120C066 (A.K.).

\bibliography{refs}

\begin{thebibliography}{30}%
\makeatletter
\providecommand \@ifxundefined [1]{%
 \@ifx{#1\undefined}
}%
\providecommand \@ifnum [1]{%
 \ifnum #1\expandafter \@firstoftwo
 \else \expandafter \@secondoftwo
 \fi
}%
\providecommand \@ifx [1]{%
 \ifx #1\expandafter \@firstoftwo
 \else \expandafter \@secondoftwo
 \fi
}%
\providecommand \natexlab [1]{#1}%
\providecommand \enquote  [1]{``#1''}%
\providecommand \bibnamefont  [1]{#1}%
\providecommand \bibfnamefont [1]{#1}%
\providecommand \citenamefont [1]{#1}%
\providecommand \href@noop [0]{\@secondoftwo}%
\providecommand \href [0]{\begingroup \@sanitize@url \@href}%
\providecommand \@href[1]{\@@startlink{#1}\@@href}%
\providecommand \@@href[1]{\endgroup#1\@@endlink}%
\providecommand \@sanitize@url [0]{\catcode `\\12\catcode `\$12\catcode
  `\&12\catcode `\#12\catcode `\^12\catcode `\_12\catcode `\%12\relax}%
\providecommand \@@startlink[1]{}%
\providecommand \@@endlink[0]{}%
\providecommand \url  [0]{\begingroup\@sanitize@url \@url }%
\providecommand \@url [1]{\endgroup\@href {#1}{\urlprefix }}%
\providecommand \urlprefix  [0]{URL }%
\providecommand \Eprint [0]{\href }%
\providecommand \doibase [0]{https://doi.org/}%
\providecommand \selectlanguage [0]{\@gobble}%
\providecommand \bibinfo  [0]{\@secondoftwo}%
\providecommand \bibfield  [0]{\@secondoftwo}%
\providecommand \translation [1]{[#1]}%
\providecommand \BibitemOpen [0]{}%
\providecommand \bibitemStop [0]{}%
\providecommand \bibitemNoStop [0]{.\EOS\space}%
\providecommand \EOS [0]{\spacefactor3000\relax}%
\providecommand \BibitemShut  [1]{\csname bibitem#1\endcsname}%
\let\auto@bib@innerbib\@empty
\bibitem [{\citenamefont {W\"achtler}\ and\ \citenamefont
  {Santos}(2016)}]{2016_PRA_Santos_Dipolar_Droplet_Theory}%
  \BibitemOpen
  \bibfield  {author} {\bibinfo {author} {\bibfnamefont {F.}~\bibnamefont
  {W\"achtler}}\ and\ \bibinfo {author} {\bibfnamefont {L.}~\bibnamefont
  {Santos}},\ }\bibfield  {title} {\bibinfo {title} {Ground-state properties
  and elementary excitations of quantum droplets in dipolar {Bose}-{Einstein}
  condensates},\ }\href {https://doi.org/10.1103/physreva.94.043618} {\bibfield
   {journal} {\bibinfo  {journal} {Phys. Rev. A}\ }\textbf {\bibinfo {volume}
  {94}},\ \bibinfo {pages} {043618} (\bibinfo {year} {2016})}\BibitemShut
  {NoStop}%
\bibitem [{\citenamefont {Petrov}(2015)}]{2015_Petrov_PRL}%
  \BibitemOpen
  \bibfield  {author} {\bibinfo {author} {\bibfnamefont {D.~S.}\ \bibnamefont
  {Petrov}},\ }\bibfield  {title} {\bibinfo {title} {Quantum mechanical
  stabilization of a collapsing {Bose}-{Bose} mixture},\ }\href
  {https://doi.org/10.1103/physrevlett.115.155302} {\bibfield  {journal}
  {\bibinfo  {journal} {Phys. Rev. Lett.}\ }\textbf {\bibinfo {volume} {115}},\
  \bibinfo {pages} {155302} (\bibinfo {year} {2015})}\BibitemShut {NoStop}%
\bibitem [{\citenamefont {Pethick}\ and\ \citenamefont
  {Smith}(2008)}]{pethick2008bose}%
  \BibitemOpen
  \bibfield  {author} {\bibinfo {author} {\bibfnamefont {C.~J.}\ \bibnamefont
  {Pethick}}\ and\ \bibinfo {author} {\bibfnamefont {H.}~\bibnamefont
  {Smith}},\ }\href {https://doi.org/10.1017/cbo9780511802850} {\emph {\bibinfo
  {title} {{Bose{\textendash}Einstein} Condensation in Dilute Gases}}}\
  (\bibinfo  {publisher} {Cambridge University Press},\ \bibinfo {year}
  {2008})\BibitemShut {NoStop}%
\bibitem [{\citenamefont {Ferrier-Barbut}\ \emph {et~al.}(2016)\citenamefont
  {Ferrier-Barbut}, \citenamefont {Kadau}, \citenamefont {Schmitt},
  \citenamefont {Wenzel},\ and\ \citenamefont
  {Pfau}}]{2016_PRL_Pfau_Dipolar_Droplet}%
  \BibitemOpen
  \bibfield  {author} {\bibinfo {author} {\bibfnamefont {I.}~\bibnamefont
  {Ferrier-Barbut}}, \bibinfo {author} {\bibfnamefont {H.}~\bibnamefont
  {Kadau}}, \bibinfo {author} {\bibfnamefont {M.}~\bibnamefont {Schmitt}},
  \bibinfo {author} {\bibfnamefont {M.}~\bibnamefont {Wenzel}},\ and\ \bibinfo
  {author} {\bibfnamefont {T.}~\bibnamefont {Pfau}},\ }\bibfield  {title}
  {\bibinfo {title} {Observation of quantum droplets in a strongly dipolar
  {Bose} gas},\ }\href {https://doi.org/10.1103/physrevlett.116.215301}
  {\bibfield  {journal} {\bibinfo  {journal} {Phys. Rev. Lett.}\ }\textbf
  {\bibinfo {volume} {116}},\ \bibinfo {pages} {215301} (\bibinfo {year}
  {2016})}\BibitemShut {NoStop}%
\bibitem [{\citenamefont {Schmitt}\ \emph {et~al.}(2016)\citenamefont
  {Schmitt}, \citenamefont {Wenzel}, \citenamefont {B\"ottcher}, \citenamefont
  {Ferrier-Barbut},\ and\ \citenamefont {Pfau}}]{2016_Nature_Pfau}%
  \BibitemOpen
  \bibfield  {author} {\bibinfo {author} {\bibfnamefont {M.}~\bibnamefont
  {Schmitt}}, \bibinfo {author} {\bibfnamefont {M.}~\bibnamefont {Wenzel}},
  \bibinfo {author} {\bibfnamefont {F.}~\bibnamefont {B\"ottcher}}, \bibinfo
  {author} {\bibfnamefont {I.}~\bibnamefont {Ferrier-Barbut}},\ and\ \bibinfo
  {author} {\bibfnamefont {T.}~\bibnamefont {Pfau}},\ }\bibfield  {title}
  {\bibinfo {title} {Self-bound droplets of a dilute magnetic quantum liquid},\
  }\href {https://doi.org/10.1038/nature20126} {\bibfield  {journal} {\bibinfo
  {journal} {Nature}\ }\textbf {\bibinfo {volume} {539}},\ \bibinfo {pages}
  {259} (\bibinfo {year} {2016})}\BibitemShut {NoStop}%
\bibitem [{\citenamefont {Macia}\ \emph {et~al.}(2016)\citenamefont {Macia},
  \citenamefont {S\'anchez-Baena}, \citenamefont {Boronat},\ and\ \citenamefont
  {Mazzanti}}]{2016_PRL_Mazzanti}%
  \BibitemOpen
  \bibfield  {author} {\bibinfo {author} {\bibfnamefont {A.}~\bibnamefont
  {Macia}}, \bibinfo {author} {\bibfnamefont {J.}~\bibnamefont
  {S\'anchez-Baena}}, \bibinfo {author} {\bibfnamefont {J.}~\bibnamefont
  {Boronat}},\ and\ \bibinfo {author} {\bibfnamefont {F.}~\bibnamefont
  {Mazzanti}},\ }\bibfield  {title} {\bibinfo {title} {Droplets of trapped
  quantum dipolar bosons},\ }\href
  {https://doi.org/10.1103/physrevlett.117.205301} {\bibfield  {journal}
  {\bibinfo  {journal} {Phys. Rev. Lett.}\ }\textbf {\bibinfo {volume} {117}},\
  \bibinfo {pages} {205301} (\bibinfo {year} {2016})}\BibitemShut {NoStop}%
\bibitem [{\citenamefont {Baillie}\ \emph {et~al.}(2016)\citenamefont
  {Baillie}, \citenamefont {Wilson}, \citenamefont {Bisset},\ and\
  \citenamefont {Blakie}}]{2016_PRA_Blakie_Dipolar_Droplet}%
  \BibitemOpen
  \bibfield  {author} {\bibinfo {author} {\bibfnamefont {D.}~\bibnamefont
  {Baillie}}, \bibinfo {author} {\bibfnamefont {R.~M.}\ \bibnamefont {Wilson}},
  \bibinfo {author} {\bibfnamefont {R.~N.}\ \bibnamefont {Bisset}},\ and\
  \bibinfo {author} {\bibfnamefont {P.~B.}\ \bibnamefont {Blakie}},\ }\bibfield
   {title} {\bibinfo {title} {Self-bound dipolar droplet: {A} localized matter
  wave in free space},\ }\href {https://doi.org/10.1103/physreva.94.021602}
  {\bibfield  {journal} {\bibinfo  {journal} {Phys. Rev. A}\ }\textbf {\bibinfo
  {volume} {94}},\ \bibinfo {pages} {021602} (\bibinfo {year}
  {2016})}\BibitemShut {NoStop}%
\bibitem [{\citenamefont {Cappellaro}\ \emph {et~al.}(2017)\citenamefont
  {Cappellaro}, \citenamefont {Macr{\`\i}}, \citenamefont {Bertacco},\ and\
  \citenamefont {Salasnich}}]{2017_Nature_Salasnisch_RabiCoupled}%
  \BibitemOpen
  \bibfield  {author} {\bibinfo {author} {\bibfnamefont {A.}~\bibnamefont
  {Cappellaro}}, \bibinfo {author} {\bibfnamefont {T.}~\bibnamefont
  {Macr{\`\i}}}, \bibinfo {author} {\bibfnamefont {G.~F.}\ \bibnamefont
  {Bertacco}},\ and\ \bibinfo {author} {\bibfnamefont {L.}~\bibnamefont
  {Salasnich}},\ }\bibfield  {title} {\bibinfo {title} {Equation of state and
  self-bound droplet in {Rabi}-coupled {Bose} mixtures},\ }\href
  {https://doi.org/10.1038/s41598-017-13647-y} {\bibfield  {journal} {\bibinfo
  {journal} {Sci. Rep.}\ }\textbf {\bibinfo {volume} {7}},\ \bibinfo {pages}
  {1} (\bibinfo {year} {2017})}\BibitemShut {NoStop}%
\bibitem [{\citenamefont {Cabrera}\ \emph {et~al.}(2018)\citenamefont
  {Cabrera}, \citenamefont {Tanzi}, \citenamefont {Sanz}, \citenamefont
  {Naylor}, \citenamefont {Thomas}, \citenamefont {Cheiney},\ and\
  \citenamefont {Tarruell}}]{2018_Science_Tarruel_Mixture_Droplet}%
  \BibitemOpen
  \bibfield  {author} {\bibinfo {author} {\bibfnamefont {C.~R.}\ \bibnamefont
  {Cabrera}}, \bibinfo {author} {\bibfnamefont {L.}~\bibnamefont {Tanzi}},
  \bibinfo {author} {\bibfnamefont {J.}~\bibnamefont {Sanz}}, \bibinfo {author}
  {\bibfnamefont {B.}~\bibnamefont {Naylor}}, \bibinfo {author} {\bibfnamefont
  {P.}~\bibnamefont {Thomas}}, \bibinfo {author} {\bibfnamefont
  {P.}~\bibnamefont {Cheiney}},\ and\ \bibinfo {author} {\bibfnamefont
  {L.}~\bibnamefont {Tarruell}},\ }\bibfield  {title} {\bibinfo {title}
  {Quantum liquid droplets in a mixture of {Bose}-{Einstein} condensates},\
  }\href {https://doi.org/10.1126/science.aao5686} {\bibfield  {journal}
  {\bibinfo  {journal} {Science}\ }\textbf {\bibinfo {volume} {359}},\ \bibinfo
  {pages} {301} (\bibinfo {year} {2018})}\BibitemShut {NoStop}%
\bibitem [{\citenamefont {Semeghini}\ \emph {et~al.}(2018)\citenamefont
  {Semeghini}, \citenamefont {Ferioli}, \citenamefont {Masi}, \citenamefont
  {Mazzinghi}, \citenamefont {Wolswijk}, \citenamefont {Minardi}, \citenamefont
  {Modugno}, \citenamefont {Modugno}, \citenamefont {Inguscio},\ and\
  \citenamefont {Fattori}}]{2018_RPL_Modugno_Mixture_Droplet}%
  \BibitemOpen
  \bibfield  {author} {\bibinfo {author} {\bibfnamefont {G.}~\bibnamefont
  {Semeghini}}, \bibinfo {author} {\bibfnamefont {G.}~\bibnamefont {Ferioli}},
  \bibinfo {author} {\bibfnamefont {L.}~\bibnamefont {Masi}}, \bibinfo {author}
  {\bibfnamefont {C.}~\bibnamefont {Mazzinghi}}, \bibinfo {author}
  {\bibfnamefont {L.}~\bibnamefont {Wolswijk}}, \bibinfo {author}
  {\bibfnamefont {F.}~\bibnamefont {Minardi}}, \bibinfo {author} {\bibfnamefont
  {M.}~\bibnamefont {Modugno}}, \bibinfo {author} {\bibfnamefont
  {G.}~\bibnamefont {Modugno}}, \bibinfo {author} {\bibfnamefont
  {M.}~\bibnamefont {Inguscio}},\ and\ \bibinfo {author} {\bibfnamefont
  {M.}~\bibnamefont {Fattori}},\ }\bibfield  {title} {\bibinfo {title}
  {Self-bound quantum droplets of atomic mixtures in free space},\ }\href
  {https://doi.org/10.1103/physrevlett.120.235301} {\bibfield  {journal}
  {\bibinfo  {journal} {Phys. Rev. Lett.}\ }\textbf {\bibinfo {volume} {120}},\
  \bibinfo {pages} {235301} (\bibinfo {year} {2018})}\BibitemShut {NoStop}%
\bibitem [{\citenamefont {Aybar}\ and\ \citenamefont
  {Oktel}(2019)}]{2019_PRA_Oktel}%
  \BibitemOpen
  \bibfield  {author} {\bibinfo {author} {\bibfnamefont {E.}~\bibnamefont
  {Aybar}}\ and\ \bibinfo {author} {\bibfnamefont {M.~O.}\ \bibnamefont
  {Oktel}},\ }\bibfield  {title} {\bibinfo {title} {Temperature-dependent
  density profiles of dipolar droplets},\ }\href
  {https://doi.org/10.1103/physreva.99.013620} {\bibfield  {journal} {\bibinfo
  {journal} {Phys. Rev. A}\ }\textbf {\bibinfo {volume} {99}},\ \bibinfo
  {pages} {013620} (\bibinfo {year} {2019})}\BibitemShut {NoStop}%
\bibitem [{\citenamefont {S\'anchez-Baena}\ \emph {et~al.}(2020)\citenamefont
  {S\'anchez-Baena}, \citenamefont {Boronat},\ and\ \citenamefont
  {Mazzanti}}]{2020_PRA_Mazzanti_SpinOrbitDroplet}%
  \BibitemOpen
  \bibfield  {author} {\bibinfo {author} {\bibfnamefont {J.}~\bibnamefont
  {S\'anchez-Baena}}, \bibinfo {author} {\bibfnamefont {J.}~\bibnamefont
  {Boronat}},\ and\ \bibinfo {author} {\bibfnamefont {F.}~\bibnamefont
  {Mazzanti}},\ }\bibfield  {title} {\bibinfo {title} {Supersolid striped
  droplets in a {Raman} spin-orbit-coupled system},\ }\href
  {https://doi.org/10.1103/physreva.102.053308} {\bibfield  {journal} {\bibinfo
   {journal} {Phys. Rev. A}\ }\textbf {\bibinfo {volume} {102}},\ \bibinfo
  {pages} {053308} (\bibinfo {year} {2020})}\BibitemShut {NoStop}%
\bibitem [{\citenamefont {Wilson}\ \emph {et~al.}(2021)\citenamefont {Wilson},
  \citenamefont {Guttridge}, \citenamefont {Segal},\ and\ \citenamefont
  {Cornish}}]{2021_PRA_Cornish_Mixture_Droplet}%
  \BibitemOpen
  \bibfield  {author} {\bibinfo {author} {\bibfnamefont {K.~E.}\ \bibnamefont
  {Wilson}}, \bibinfo {author} {\bibfnamefont {A.}~\bibnamefont {Guttridge}},
  \bibinfo {author} {\bibfnamefont {J.}~\bibnamefont {Segal}},\ and\ \bibinfo
  {author} {\bibfnamefont {S.~L.}\ \bibnamefont {Cornish}},\ }\bibfield
  {title} {\bibinfo {title} {Quantum degenerate mixtures of {Cs} and {Yb}},\
  }\href {https://doi.org/10.1103/physreva.103.033306} {\bibfield  {journal}
  {\bibinfo  {journal} {Phys. Rev. A}\ }\textbf {\bibinfo {volume} {103}},\
  \bibinfo {pages} {033306} (\bibinfo {year} {2021})}\BibitemShut {NoStop}%
\bibitem [{\citenamefont {Ma}\ \emph {et~al.}(2021)\citenamefont {Ma},
  \citenamefont {Peng},\ and\ \citenamefont
  {Cui}}]{2021_Arxiv_Cui_Borromean_Droplet}%
  \BibitemOpen
  \bibfield  {author} {\bibinfo {author} {\bibfnamefont {Y.}~\bibnamefont
  {Ma}}, \bibinfo {author} {\bibfnamefont {C.}~\bibnamefont {Peng}},\ and\
  \bibinfo {author} {\bibfnamefont {X.}~\bibnamefont {Cui}},\ }\bibfield
  {title} {\bibinfo {title} {{Borromean} droplet in three-component ultracold
  {Bose} gases},\ }\href {https://doi.org/10.1103/physrevlett.127.043002}
  {\bibfield  {journal} {\bibinfo  {journal} {Phys. Rev. Lett.}\ }\textbf
  {\bibinfo {volume} {127}} (\bibinfo {year} {2021})}\BibitemShut {NoStop}%
\bibitem [{\citenamefont {Bisset}\ \emph {et~al.}(2021)\citenamefont {Bisset},
  \citenamefont {Ardila},\ and\ \citenamefont
  {Santos}}]{2021_PRL_Santos_Dipolar_Mixture}%
  \BibitemOpen
  \bibfield  {author} {\bibinfo {author} {\bibfnamefont {R.~N.}\ \bibnamefont
  {Bisset}}, \bibinfo {author} {\bibfnamefont {L.~A. P.~n.}\ \bibnamefont
  {Ardila}},\ and\ \bibinfo {author} {\bibfnamefont {L.}~\bibnamefont
  {Santos}},\ }\bibfield  {title} {\bibinfo {title} {Quantum droplets of
  dipolar mixtures},\ }\href {https://doi.org/10.1103/physrevlett.126.025301}
  {\bibfield  {journal} {\bibinfo  {journal} {Phys. Rev. Lett.}\ }\textbf
  {\bibinfo {volume} {126}},\ \bibinfo {pages} {025301} (\bibinfo {year}
  {2021})}\BibitemShut {NoStop}%
\bibitem [{\citenamefont {Yo\u{g}urt}\ \emph {et~al.}(2022)\citenamefont
  {Yo\u{g}urt}, \citenamefont {Kele\c{s}},\ and\ \citenamefont
  {Oktel}}]{2022_PRA_Yogurt_Spinor}%
  \BibitemOpen
  \bibfield  {author} {\bibinfo {author} {\bibfnamefont {T.~A.}\ \bibnamefont
  {Yo\u{g}urt}}, \bibinfo {author} {\bibfnamefont {A.}~\bibnamefont
  {Kele\c{s}}},\ and\ \bibinfo {author} {\bibfnamefont {M.~O.}\ \bibnamefont
  {Oktel}},\ }\bibfield  {title} {\bibinfo {title} {Spinor boson droplets
  stabilized by spin fluctuations},\ }\href
  {https://doi.org/10.1103/physreva.105.043309} {\bibfield  {journal} {\bibinfo
   {journal} {Phys. Rev. A}\ }\textbf {\bibinfo {volume} {105}},\ \bibinfo
  {pages} {043309} (\bibinfo {year} {2022})}\BibitemShut {NoStop}%
\bibitem [{\citenamefont
  {Chiquillo}(2019)}]{2019_PRA_Chiquillo_Rabicoupled_lowdimensional}%
  \BibitemOpen
  \bibfield  {author} {\bibinfo {author} {\bibfnamefont {E.}~\bibnamefont
  {Chiquillo}},\ }\bibfield  {title} {\bibinfo {title} {Low-dimensional
  self-bound quantum {Rabi}-coupled bosonic droplets},\ }\href
  {https://doi.org/10.1103/physreva.99.051601} {\bibfield  {journal} {\bibinfo
  {journal} {Phys. Rev. A}\ }\textbf {\bibinfo {volume} {99}},\ \bibinfo
  {pages} {051601} (\bibinfo {year} {2019})}\BibitemShut {NoStop}%
\bibitem [{\citenamefont {Lavoine}\ \emph {et~al.}(2021)\citenamefont
  {Lavoine}, \citenamefont {Hammond}, \citenamefont {Recati}, \citenamefont
  {Petrov},\ and\ \citenamefont {Bourdel}}]{2021_PRL_Bourdel_Rabicoupled_BMF}%
  \BibitemOpen
  \bibfield  {author} {\bibinfo {author} {\bibfnamefont {L.}~\bibnamefont
  {Lavoine}}, \bibinfo {author} {\bibfnamefont {A.}~\bibnamefont {Hammond}},
  \bibinfo {author} {\bibfnamefont {A.}~\bibnamefont {Recati}}, \bibinfo
  {author} {\bibfnamefont {D.~S.}\ \bibnamefont {Petrov}},\ and\ \bibinfo
  {author} {\bibfnamefont {T.}~\bibnamefont {Bourdel}},\ }\bibfield  {title}
  {\bibinfo {title} {Beyond-mean-field effects in {Rabi}-coupled two-component
  {Bose}-{Einstein} condensate},\ }\href
  {https://doi.org/10.1103/physrevlett.127.203402} {\bibfield  {journal}
  {\bibinfo  {journal} {Phys. Rev. Lett.}\ }\textbf {\bibinfo {volume} {127}},\
  \bibinfo {pages} {203402} (\bibinfo {year} {2021})}\BibitemShut {NoStop}%
\bibitem [{\citenamefont {Hammond}\ \emph {et~al.}(2022)\citenamefont
  {Hammond}, \citenamefont {Lavoine},\ and\ \citenamefont
  {Bourdel}}]{2022_PRL_Bourdel_Rabicoupled_Threebody}%
  \BibitemOpen
  \bibfield  {author} {\bibinfo {author} {\bibfnamefont {A.}~\bibnamefont
  {Hammond}}, \bibinfo {author} {\bibfnamefont {L.}~\bibnamefont {Lavoine}},\
  and\ \bibinfo {author} {\bibfnamefont {T.}~\bibnamefont {Bourdel}},\
  }\bibfield  {title} {\bibinfo {title} {Tunable three-body interactions in
  driven two-component {Bose}-{Einstein} condensates},\ }\href
  {https://doi.org/10.1103/physrevlett.128.083401} {\bibfield  {journal}
  {\bibinfo  {journal} {Phys. Rev. Lett.}\ }\textbf {\bibinfo {volume} {128}},\
  \bibinfo {pages} {083401} (\bibinfo {year} {2022})}\BibitemShut {NoStop}%
\bibitem [{\citenamefont {Kawaguchi}\ and\ \citenamefont
  {Ueda}(2012)}]{2012_PR_Ueda_Spinor_BEC}%
  \BibitemOpen
  \bibfield  {author} {\bibinfo {author} {\bibfnamefont {Y.}~\bibnamefont
  {Kawaguchi}}\ and\ \bibinfo {author} {\bibfnamefont {M.}~\bibnamefont
  {Ueda}},\ }\bibfield  {title} {\bibinfo {title} {Spinor
  {Bose{\textendash}Einstein} condensates},\ }\href
  {https://doi.org/10.1016/j.physrep.2012.07.005} {\bibfield  {journal}
  {\bibinfo  {journal} {Phys. Rep.}\ }\textbf {\bibinfo {volume} {520}},\
  \bibinfo {pages} {253} (\bibinfo {year} {2012})}\BibitemShut {NoStop}%
\bibitem [{\citenamefont {Stamper-Kurn}\ and\ \citenamefont
  {Ueda}(2013)}]{2013_RMP_Spinor_Kurn_Ueda}%
  \BibitemOpen
  \bibfield  {author} {\bibinfo {author} {\bibfnamefont {D.~M.}\ \bibnamefont
  {Stamper-Kurn}}\ and\ \bibinfo {author} {\bibfnamefont {M.}~\bibnamefont
  {Ueda}},\ }\bibfield  {title} {\bibinfo {title} {Spinor {Bose} gases:
  {Symmetries,} magnetism, and quantum dynamics},\ }\href
  {https://doi.org/10.1103/revmodphys.85.1191} {\bibfield  {journal} {\bibinfo
  {journal} {Rev. Mod. Phys.}\ }\textbf {\bibinfo {volume} {85}},\ \bibinfo
  {pages} {1191} (\bibinfo {year} {2013})}\BibitemShut {NoStop}%
\bibitem [{\citenamefont {Guzman}\ \emph {et~al.}(2011)\citenamefont {Guzman},
  \citenamefont {Jo}, \citenamefont {Wenz}, \citenamefont {Murch},
  \citenamefont {Thomas},\ and\ \citenamefont
  {Stamper-Kurn}}]{2011_PRA_Kurn_Spinor}%
  \BibitemOpen
  \bibfield  {author} {\bibinfo {author} {\bibfnamefont {J.}~\bibnamefont
  {Guzman}}, \bibinfo {author} {\bibfnamefont {G.-B.}\ \bibnamefont {Jo}},
  \bibinfo {author} {\bibfnamefont {A.~N.}\ \bibnamefont {Wenz}}, \bibinfo
  {author} {\bibfnamefont {K.~W.}\ \bibnamefont {Murch}}, \bibinfo {author}
  {\bibfnamefont {C.~K.}\ \bibnamefont {Thomas}},\ and\ \bibinfo {author}
  {\bibfnamefont {D.~M.}\ \bibnamefont {Stamper-Kurn}},\ }\bibfield  {title}
  {\bibinfo {title} {Long-time-scale dynamics of spin textures in a degenerate
  {F=1} $^{87}${Rb} spinor {Bose} gas},\ }\href
  {https://doi.org/10.1103/physreva.84.063625} {\bibfield  {journal} {\bibinfo
  {journal} {Phys. Rev. A}\ }\textbf {\bibinfo {volume} {84}},\ \bibinfo
  {pages} {063625} (\bibinfo {year} {2011})}\BibitemShut {NoStop}%
\bibitem [{\citenamefont {Uchino}\ \emph {et~al.}(2010)\citenamefont {Uchino},
  \citenamefont {Kobayashi},\ and\ \citenamefont
  {Ueda}}]{2010_PRA_Ueda_Spinor_LHY}%
  \BibitemOpen
  \bibfield  {author} {\bibinfo {author} {\bibfnamefont {S.}~\bibnamefont
  {Uchino}}, \bibinfo {author} {\bibfnamefont {M.}~\bibnamefont {Kobayashi}},\
  and\ \bibinfo {author} {\bibfnamefont {M.}~\bibnamefont {Ueda}},\ }\bibfield
  {title} {\bibinfo {title} {Bogoliubov theory and {Lee}-{Huang}-{Yang}
  corrections in spin-1 and spin-2 {Bose}-{Einstein} condensates in the
  presence of the quadratic {Zeeman} effect},\ }\href
  {https://doi.org/10.1103/physreva.81.063632} {\bibfield  {journal} {\bibinfo
  {journal} {Phys. Rev. A}\ }\textbf {\bibinfo {volume} {81}},\ \bibinfo
  {pages} {063632} (\bibinfo {year} {2010})}\BibitemShut {NoStop}%
\bibitem [{\citenamefont {Stenger}\ \emph {et~al.}(1998)\citenamefont
  {Stenger}, \citenamefont {Inouye}, \citenamefont {Stamper-Kurn},
  \citenamefont {Miesner}, \citenamefont {Chikkatur},\ and\ \citenamefont
  {Ketterle}}]{1998_Nature_Ketterle_Spinor}%
  \BibitemOpen
  \bibfield  {author} {\bibinfo {author} {\bibfnamefont {J.}~\bibnamefont
  {Stenger}}, \bibinfo {author} {\bibfnamefont {S.}~\bibnamefont {Inouye}},
  \bibinfo {author} {\bibfnamefont {D.~M.}\ \bibnamefont {Stamper-Kurn}},
  \bibinfo {author} {\bibfnamefont {H.-J.}\ \bibnamefont {Miesner}}, \bibinfo
  {author} {\bibfnamefont {A.~P.}\ \bibnamefont {Chikkatur}},\ and\ \bibinfo
  {author} {\bibfnamefont {W.}~\bibnamefont {Ketterle}},\ }\bibfield  {title}
  {\bibinfo {title} {Spin domains in ground-state {Bose{\textendash}Einstein}
  condensates},\ }\href {https://doi.org/10.1038/24567} {\bibfield  {journal}
  {\bibinfo  {journal} {Nature}\ }\textbf {\bibinfo {volume} {396}},\ \bibinfo
  {pages} {345} (\bibinfo {year} {1998})}\BibitemShut {NoStop}%
\bibitem [{\citenamefont {Chang}\ \emph {et~al.}(2004)\citenamefont {Chang},
  \citenamefont {Hamley}, \citenamefont {Barrett}, \citenamefont {Sauer},
  \citenamefont {Fortier}, \citenamefont {Zhang}, \citenamefont {You},\ and\
  \citenamefont {Chapman}}]{2004_PRL_Chapman_spinor}%
  \BibitemOpen
  \bibfield  {author} {\bibinfo {author} {\bibfnamefont {M.-S.}\ \bibnamefont
  {Chang}}, \bibinfo {author} {\bibfnamefont {C.~D.}\ \bibnamefont {Hamley}},
  \bibinfo {author} {\bibfnamefont {M.~D.}\ \bibnamefont {Barrett}}, \bibinfo
  {author} {\bibfnamefont {J.~A.}\ \bibnamefont {Sauer}}, \bibinfo {author}
  {\bibfnamefont {K.~M.}\ \bibnamefont {Fortier}}, \bibinfo {author}
  {\bibfnamefont {W.}~\bibnamefont {Zhang}}, \bibinfo {author} {\bibfnamefont
  {L.}~\bibnamefont {You}},\ and\ \bibinfo {author} {\bibfnamefont {M.~S.}\
  \bibnamefont {Chapman}},\ }\bibfield  {title} {\bibinfo {title} {Observation
  of spinor dynamics in optically trapped $^{87}${Rb} {Bose}-{Einstein}
  condensates},\ }\href {https://doi.org/10.1103/physrevlett.92.140403}
  {\bibfield  {journal} {\bibinfo  {journal} {Phys. Rev. Lett.}\ }\textbf
  {\bibinfo {volume} {92}},\ \bibinfo {pages} {140403} (\bibinfo {year}
  {2004})}\BibitemShut {NoStop}%
\bibitem [{\citenamefont {Black}\ \emph {et~al.}(2007)\citenamefont {Black},
  \citenamefont {Gomez}, \citenamefont {Turner}, \citenamefont {Jung},\ and\
  \citenamefont {Lett}}]{2007_PRL_Let_RF_Quadratic}%
  \BibitemOpen
  \bibfield  {author} {\bibinfo {author} {\bibfnamefont {A.}~\bibnamefont
  {Black}}, \bibinfo {author} {\bibfnamefont {E.}~\bibnamefont {Gomez}},
  \bibinfo {author} {\bibfnamefont {L.}~\bibnamefont {Turner}}, \bibinfo
  {author} {\bibfnamefont {S.}~\bibnamefont {Jung}},\ and\ \bibinfo {author}
  {\bibfnamefont {P.}~\bibnamefont {Lett}},\ }\bibfield  {title} {\bibinfo
  {title} {Spinor dynamics in an antiferromagnetic spin-1 condensate},\ }\href
  {https://doi.org/10.1103/physrevlett.99.070403} {\bibfield  {journal}
  {\bibinfo  {journal} {Phys. Rev. Lett.}\ }\textbf {\bibinfo {volume} {99}},\
  \bibinfo {pages} {070403} (\bibinfo {year} {2007})}\BibitemShut {NoStop}%
\bibitem [{\citenamefont {Huh}\ \emph {et~al.}(2020)\citenamefont {Huh},
  \citenamefont {Kim}, \citenamefont {Kwon},\ and\ \citenamefont
  {Choi}}]{2020_PRR_Choi}%
  \BibitemOpen
  \bibfield  {author} {\bibinfo {author} {\bibfnamefont {S.}~\bibnamefont
  {Huh}}, \bibinfo {author} {\bibfnamefont {K.}~\bibnamefont {Kim}}, \bibinfo
  {author} {\bibfnamefont {K.}~\bibnamefont {Kwon}},\ and\ \bibinfo {author}
  {\bibfnamefont {J.-y.}\ \bibnamefont {Choi}},\ }\bibfield  {title} {\bibinfo
  {title} {Observation of a strongly ferromagnetic spinor {Bose}-{Einstein}
  condensate},\ }\href {https://doi.org/10.1103/physrevresearch.2.033471}
  {\bibfield  {journal} {\bibinfo  {journal} {Phys. Rev. Research}\ }\textbf
  {\bibinfo {volume} {2}},\ \bibinfo {pages} {033471} (\bibinfo {year}
  {2020})}\BibitemShut {NoStop}%
\bibitem [{\citenamefont {Nicholson}\ \emph {et~al.}(2015)\citenamefont
  {Nicholson}, \citenamefont {Blatt}, \citenamefont {Bloom}, \citenamefont
  {Williams}, \citenamefont {Thomsen}, \citenamefont {Ye},\ and\ \citenamefont
  {Julienne}}]{2015_PRA_Optical_Feschbach_Resonance}%
  \BibitemOpen
  \bibfield  {author} {\bibinfo {author} {\bibfnamefont {T.~L.}\ \bibnamefont
  {Nicholson}}, \bibinfo {author} {\bibfnamefont {S.}~\bibnamefont {Blatt}},
  \bibinfo {author} {\bibfnamefont {B.~J.}\ \bibnamefont {Bloom}}, \bibinfo
  {author} {\bibfnamefont {J.~R.}\ \bibnamefont {Williams}}, \bibinfo {author}
  {\bibfnamefont {J.~W.}\ \bibnamefont {Thomsen}}, \bibinfo {author}
  {\bibfnamefont {J.}~\bibnamefont {Ye}},\ and\ \bibinfo {author}
  {\bibfnamefont {P.~S.}\ \bibnamefont {Julienne}},\ }\bibfield  {title}
  {\bibinfo {title} {Optical {Feshbach} resonances: {Field-dressed} theory and
  comparison with experiments},\ }\href
  {https://doi.org/10.1103/physreva.92.022709} {\bibfield  {journal} {\bibinfo
  {journal} {Phys. Rev. A}\ }\textbf {\bibinfo {volume} {92}},\ \bibinfo
  {pages} {022709} (\bibinfo {year} {2015})}\BibitemShut {NoStop}%
\bibitem [{\citenamefont {Thomas}\ \emph {et~al.}(2018)\citenamefont {Thomas},
  \citenamefont {Lippe}, \citenamefont {Eichert},\ and\ \citenamefont
  {Ott}}]{2018_Nature_Ott_Optical_Feshbach_Resonance}%
  \BibitemOpen
  \bibfield  {author} {\bibinfo {author} {\bibfnamefont {O.}~\bibnamefont
  {Thomas}}, \bibinfo {author} {\bibfnamefont {C.}~\bibnamefont {Lippe}},
  \bibinfo {author} {\bibfnamefont {T.}~\bibnamefont {Eichert}},\ and\ \bibinfo
  {author} {\bibfnamefont {H.}~\bibnamefont {Ott}},\ }\bibfield  {title}
  {\bibinfo {title} {Experimental realization of a {Rydberg} optical {Feshbach}
  resonance in a quantum many-body system},\ }\href
  {https://doi.org/10.1038/s41467-018-04684-w} {\bibfield  {journal} {\bibinfo
  {journal} {Nat. Commun.}\ }\textbf {\bibinfo {volume} {9}},\ \bibinfo {pages}
  {1} (\bibinfo {year} {2018})}\BibitemShut {NoStop}%
\bibitem [{\citenamefont {Gu}\ and\ \citenamefont
  {Cui}(2022)}]{2022_Arxiv_Cui_Polarized_RabiCoupled}%
  \BibitemOpen
  \bibfield  {author} {\bibinfo {author} {\bibfnamefont {Q.}~\bibnamefont
  {Gu}}\ and\ \bibinfo {author} {\bibfnamefont {X.}~\bibnamefont {Cui}},\
  }\bibfield  {title} {\bibinfo {title} {Liquid-gas coexistence in binary
  {Bose-Einstein} condensates},\ }\href {https://arxiv.org/abs/2209.10019}
  {\bibfield  {journal} {\bibinfo  {journal} {arXiv preprint arXiv:2209.10019}\
  } (\bibinfo {year} {2022})}\BibitemShut {NoStop}%
\end{thebibliography}%


%

\end{document}